\documentclass[aps,prd,twocolumn,showpacs,showkeys,10pt]{revtex4-1}

\usepackage[english]{babel}
\usepackage[utf8]{inputenc}   
\usepackage[T1]{fontenc} 
\usepackage{amsmath, amsfonts, amssymb, amsthm, mathrsfs, bm, mathtools}
\usepackage{yhmath}
\usepackage{braket}
\usepackage{enumerate}
\usepackage{mdwlist}
\usepackage{tensor}
\usepackage{slashed} 
\usepackage{empheq}
\usepackage[makeroom]{cancel}
\usepackage{simplewick} 

\usepackage[breaklinks]{hyperref}
\usepackage{xcolor}
\hypersetup{
    colorlinks,
    linkcolor={red!80!black},
    citecolor={green!70!black},
    urlcolor={blue!80!black}
}
\colorlet{dkgreen}{green!50!black}

\usepackage{tikz}

\usepackage[]{natbib}

\renewcommand{\vec}{\mathbf}
\renewcommand{\Pi}{\varPi}
\newcommand*{\diff}{\mathop{}\!\mathrm{d}}
\newcommand*\Diff[1]{\mathop{}\!\mathrm{d}^#1}
\newcommand*\fundiff[1]{\mathop{}\!\mathcal{D}#1\,}

\DeclareMathOperator{\Tr}{Tr}
\DeclareMathOperator{\tr}{tr}

\newcommand{\Rnum}[1]{%
  \textup{\uppercase\expandafter{\romannumeral#1}}%
}

\begin{document}

\title{Effective mesonic theory for the 't~Hooft model on the lattice}
\date{\today}
\author{Sergio Caracciolo}
\email{sergio.caracciolo@mi.infn.it}
\author{Mauro Pastore}
\email{mauro.pastore@unimi.it}
\affiliation{Dipartimento di Fisica, Universit\`a degli Studi di Milano}
\affiliation{INFN, Via Celoria 16, I-20133 Milan, Italy}

\begin{abstract}
We apply to a lattice version of the 't~Hooft model, QCD in two space-time dimensions for large number of colours, a method recently proposed to obtain an effective mesonic action starting from the fundamental, fermionic one. The idea is to pass from a canonical, operatorial representation, where the low-energy states have a direct physical interpretation in terms of a Bogoliubov vacuum and its corresponding quasiparticle excitations, to a functional, path integral representation, via the formalism of the transfer matrix. In this way we obtain a lattice effective theory for mesons in a self-consistent setting. We also verify that well-known results from other different approaches are reproduced in the continuum limit.
\end{abstract}

\maketitle


\section{\label{sec:intro}Introduction}

Since its formulation by 't~Hooft in 1974, QCD in two space-time dimensions in the limit for large number of colours (in the following: \emph{'t~Hooft model}, or QCD$_2$) has been very successful as a toy model for strong interactions. Despite its relative simplicity, it exhibits highly nontrivial properties (such as confinement, spontaneous breakdown of chiral symmetry, a bound-state spectrum of excitations, ...) which makes it a per se worthwhile study subject and an ideal benchmark for new ideas. In addition to the fundamental works~\cite{tHooft:1974pnl,bars:1977ud,li:1986gf,li:1987hx}, the interested reader can find clues of the richness of its features in reviews like~\cite{abdalla:1995dm}.

The aim of the present work is to apply to QCD$_2$ a very general formalism introduced in many-body theory~\cite{palumbo:2007nn,palumbo:2005pv} and extended to relativistic field theories in~\cite{caracciolo:2006wc,caracciolo:2008ag,caracciolo:2010xe,caracciolo:2010rm,caracciolo:2011aa} to obtain a theory of effective mesonic excitations above a non-perturbative vacuum state, starting from the underlying theory of fundamental fermions (in this case, quarks). The method consists in two main ingredients:
\begin{itemize}
\item The canonical, operator formulation of the model, where the fermionic Fock space of states is explicitly built acting on the vacuum with creation and annihilation operators. Because of confinement, in the low energy phase of the model a condensation of the fermions into composites occurs, with some analogies with BCS-like theories~\cite{bogolyubov:1962zz}: there, the onset of superconductivity is justified by the existence of an energy gap between the ground state of non-interactive conduction electrons and a state, obtained from the first via a Bogoliubov transformation and a variational principle, where the fermions are coupled in weakly bounded states, the Cooper pairs. Here, a similar mechanism can be invoked to describe the formation of a non-perturbative vacuum where quarks and antiquarks form mesonic condensates.
\item The holomorphic representation~\cite{faddeev:mio, FS} of the functional integral defining the partition function of the model. In order to have a mathematically well defined setting we use the Euclidean formulation of the model regularized on a lattice, where the dynamics is controlled by the  \emph{transfer matrix} which maps the Fock spaces defined at different time slices one into the other. The partition function is obtained evaluating the trace of the Boltzmann factor over the fermionic Fock space on a basis of canonical coherent states. Moreover, under the assumption of \emph{composite boson dominance}~\cite{palumbo:2007nn,palumbo:2005pv,caracciolo:2006wc}, according to which the low-energy spectrum of the theory is dominated by bosonic modes, a projection of the transfer matrix over states built as mesonic excitations of the non-perturbative vacuum  gives an approximate functional representation of the partition function in terms of an action for composite bosonic fields.
\end{itemize}
Along these lines we can pass from canonical to functional formalism and back, taking the best from each: physical intuition and a non-perturbative approach from the first, a full relativistic field theory formulation from the latter. In the 't~Hooft model we end up with an effective action for mesons on the lattice which bosonizes the model and reproduces, in the continuum limit, results well known from perturbative~\cite{tHooft:1974pnl,bars:1977ud} or Hamiltonian canonical~\cite{bardeen:1988mh,kalashnikova:2001df} approaches.

The work is organized as follows: in section~\ref{sec:bogoliubov}, after a brief review of the general method presented in~\cite{caracciolo:2008ag}, we explain how the hypothesis of composite boson dominance can be used to obtain an effective mesonic description of a fermionic theory (subsection \ref{subsec:mesons}). In section~\ref{sec:qcd2} we present the 't~Hooft model, in Coulomb gauge and in Wilson's lattice formulation. The section~\ref{sec:qcd2_effective} is dedicated to the application of our formalism to this model until, in section~\ref{subsec:qcd2_mesons}, we derive an action for mesons in QCD$_2$ and show how it has the correct continuum limit, equation \eqref{eq:QCD2_action}. In commenting on the results, in subsections \ref{subsec:vacuum} and \ref{subsec:qcd2_mesons}, we stress similarities and differences between ours and the previous works on the subject, with a particular consideration to the ones by W. Bardeen~\cite{bardeen:1988mh} and Kalashnikova and Nefed'ev~\cite{kalashnikova:2001df}, which are the most closely related to our present approach. Finally, we draw our conclusions in section~\ref{sec:conclusion}. Some formulas and technicalities, useful to re-obtain our results, are collected in the appendices.

A preliminary report of the present work has been presented at the XIII Quark Confinement and the Hadron Spectrum conference (Maynooth, August 2018). We refer the reader to the corresponding Proceedings for a concise exposition of our reasonings and results.

\section{\label{sec:bogoliubov}Effective action with Bogoliubov transformations}
This section starts with a review of methods and concepts fully developed in reference~\cite{caracciolo:2008ag} and, unless explicitly stated, concerns fermionic lattice gauge theories in arbitrary space-time dimensions with very generic features. The last part, \ref{subsec:mesons}, presents a new point of view on the hypothesis of composite boson dominance in the spectrum of excitations of the theory.

\subsection{Transfer matrix for fermionic theories}
\label{subsec:transfer_matrix}
The quantum theory of fermions interacting with a gauge field descends, in the Euclidean  functional formalism, from the partition function
\begin{align}
\mathcal{Z} = & \int \!\fundiff U\! \fundiff \psi\! \fundiff{\bar{\psi}}\, e^{-S[\psi,\bar{\psi},U]} \\
= &  \int \!\fundiff U\! e^{-S_G[U]} \, \int \!\fundiff \psi\! \fundiff{\bar{\psi}}\,  e^{-S_F[\psi,\bar{\psi},U]}\\
= & \int \!\fundiff U\! e^{-S_G[U]} \, \mathcal{Z}_F
\end{align}
where, in the total action, we distinguished the pure gauge part $S_G$
\begin{equation}
S[\psi,\bar{\psi},U] = S_F[\psi,\bar{\psi},U] + S_G[U]
\end{equation}
and we introduced the fermionic partition function in the external gauge configuration
\begin{equation}
\mathcal{Z}_F[U] = \int \!\fundiff \psi\! \fundiff{\bar{\psi}}\,  e^{-S_F[\psi,\bar{\psi},U]}\, .\label{eq:transfer_partition_path}
\end{equation}
The fermionc measure is defined by
\begin{equation}
\fundiff \psi\! \fundiff{\bar{\psi}} =\prod_x \diff{\psi(x)} \diff{\bar{\psi}(x)}\, ,
\end{equation}
where $\diff{\psi(x)} \diff{\bar{\psi}(x)}$ denotes a Berezin integral over the Grassmann variables $\psi$, $\bar{\psi}$ defined at the site $x$, representing fermions, while $\fundiff U$ is the product of the Haar measures on all the gauge link variables which live, in the lattice formulation, in the gauge group SU($N_c$). 

The bridge between the functional and the operatorial formalism is provided by the partition function itself, written in terms of the Hamiltonian operator $\hat{H}_F$ for the fermions:
\begin{equation}
\mathcal{Z}_F = \Tr^F\, e^{- \beta \hat{H}_F},
\label{eq:transfer_partition_statistical_H}
\end{equation}
where $\Tr^F$ denotes a sum over fermionic Fock space and $\beta$ represents the total time of evolution.
Strictly speaking, on the lattice the Hamiltonian, which  is the generator of continuum time evolution, is replaced by the \emph{transfer matrix} $\hat{\mathcal{T}}_{t,t+1}$, the operator that maps the Fock space defined at time $t$ into that defined at time $t+1$, where time direction is a chosen lattice direction. Furthermore, as in our problem there is a dependence form the gauge configuration, which is time dependent, in the product appearing in
\begin{equation}
\mathcal{Z}_F = \Tr^F\, \prod_t \hat{\mathcal{T}}_{t,t+1}
\label{eq:transfer_partition_statistical_T}
\end{equation}
we must follow an ordering in time of the different factors.
If the transfer matrix is self-adjoint and strictly positive, then a lattice Hamiltonian can be defined as
\begin{equation}
\hat{H}_F = - \frac{1}{a_0}\ln \hat{\mathcal{T}}_{t,t+1} \, ,
\label{eq:transfer_Hamiltonian}
\end{equation}
with $a_0$ the lattice spacing in the time direction. The proof of the existence of such an operator for Wilson fermions in the temporal gauge $U_0 = 1$ can be found in~\cite{luescher:1976ms,creutz:1986ky,smit:1990pb,smit:lqft}. For the construction of the transfer matrix for Kogut-Susskind fermions in the flavor basis see~\cite{palumbo2002} and erratum in~\cite{palumbo2006}, while for the spin basis see~\cite{Caracciolo:155}.  In order to work in a more general gauge, we will use a slightly different method, as in~\cite{caracciolo:2008ag}. Then, the fermionic partition function~\eqref{eq:transfer_partition_path} can be written as
\begin{equation}
\mathcal{Z}_F = \Tr^F \prod_t \mathcal{J}_t\hat{\mathcal{T}}_{t,t+1} \, ,\label{eq:transfer_partition_transfer}
\end{equation}
where the operator $\hat{\mathcal{T}}_{t,t+1}$ can be put in the form
\begin{equation}
\hat{\mathcal{T}}_{t,t+1} =  \hat{T}^\dagger_t \hat{V}_t \hat{T}_{t+1}
\label{eq:transfer_matrix}
\end{equation}
with
\begin{align}
\hat{T}_{t} &= \exp \left[-\hat{u}^\dagger \mathcal{M}_t \hat{u} +  \hat{v}\mathcal{M}_t \hat{v}^\dagger \right] \exp\left[\hat{v}\mathcal{N}_t \hat{u} \right]\, ,\\
\hat{V}_{t} &= \exp \left[\hat{u}^\dagger \ln \mathcal{U}_{0,t} \hat{u} -   \hat{v}\ln \mathcal{U}^\dagger_{0,t}\hat{v}^\dagger \right]\, ,\\
\mathcal{J}_t &= \exp{\left[\Tr \left(\mathcal{M}_t + \mathcal{M}^\dagger_t \right)\right]}
\end{align}
and $\mathcal{M}_t$, $\mathcal{N}_t$, $\mathcal{U}_{0,t}$ matrices in internal (colour) and space (time enters as a label only because the external gauge configuration is time dependent) indices, with the last one defined as
\begin{equation}
[\mathcal{U}_{\mu,t}]_{n_1 n_2}\equiv  \delta_{n_1 n_2} (U_\mu)_{n_1,t} \,.
\end{equation}
Therefore, ``$\Tr$'' is a sum over these indices. The symbols $\hat{u}^\dagger$, respectively $\hat{v}^\dagger$, denote the creation operator of fermions, respectively antifermions, and carry internal and space indices, which are understood. In the {\em standard basis} the fermion field has the form
\begin{equation} 
\hat{\psi} =  \left(\begin{array}{c}\hat{u}  \\ \hat{v}^\dagger \end{array}\right)\, .
\label{eq:transfer_std_basis}
\end{equation}

The kernel of an operator can be expressed in terms of matrix elements between canonical coherent states; using formulas collected in Appendix~\ref{subsec:A_coherent} it's easy to get, from the transfer matrix~\eqref{eq:transfer_matrix}, the usual functional representation in term of anticommuting fields.

\subsection{Bogoliubov transformations}

As long as the theory exhibits \emph{confinement}, the elementary fermions created by the operators $\hat{u}^\dagger$ and $\hat{v}^\dagger$ are not present in the low-energy spectrum:  in building up the Fock space from the ``perturbative'' vacuum $\ket{0}$, defined as the state annihilated by $\hat{u}$ and $\hat{v}$, we are adopting a point of view quite awkward to describe this phase of the theory. The true (``physical'') vacuum state, defined as the state of minimum energy, is strictly \emph{below} the perturbative one in the spectrum, so that a gap is produced that makes the bound state more favorable. The fundamental tool we use to study this non-perturbative vacuum is the same as the one used in BCS theory~\cite{bardeen:1957mv} to explain the formation of Cooper pairs in the superconductive phase: a linear transformation that preserves canonical anticommutation rules is implemented on the fermion ladder operators. This \emph{Bogoliubov transformation} mixes creators and annihilators in such a way that induces a transformation in the Fock space: the state whose particle content cannot be lowered by the new annihilators is different from the starting vacuum $\ket{0}$ and depends on the parameters of the transformation. Then, the transformation is fixed by a variational principle (a request of minimal energy).

We define a linear transformation in the space of canonical operators as
\begin{equation}
\begin{aligned}
&\hat{a} = \mathcal{R}^{1/2} \left(\hat{u} - \mathcal{F}^\dagger \hat{v}^\dagger \right)\, ,\\
&\hat{b} = \left(\hat{v} +  \hat{u}^\dagger \mathcal{F}^\dagger \right) \mathring{\mathcal{R}}^{1/2}\,,
\end{aligned}
\label{eq:unitary_bogoliubov}
\end{equation}
where $\mathcal{F}$ is an arbitrary matrix  and
\begin{equation}
\mathcal{R} = \left(1+\mathcal{F}^\dagger\mathcal{F} \right)^{-1}\,, \qquad
\mathring{\mathcal{R}} = \left(1+\mathcal{F}\mathcal{F}^\dagger \right)^{-1}\,.
\label{eq:unitary_R}
\end{equation}
This transformation in \emph{canonical}, in the sense that it preserves canonical anticommutation relations:
\begin{equation}
\begin{aligned}
&\left\{\hat{a}_J,\hat{a}^\dagger_K \right\} = \delta_{JK} = \left\{\hat{b}_J,\hat{b}^\dagger_K \right\},\\
&\left\{\hat{a}_J,\hat{a}_K \right\} = \left\{\hat{b}_J,\hat{b}_K \right\} = \left\{\hat{a}_J,\hat{b}_K \right\} = \left\{\hat{a}_J,\hat{b}_K^\dagger \right\} = 0
\end{aligned}
\label{eq:unitary_CAR_ab}
\end{equation}
where capitol letters are multi-indices.
The Bogouliubov transformation mixes creation and annihilation operators, so it transforms implicitly the Fock space, starting from the new (not yet normalized) vacuum
\begin{equation}
\ket{\mathcal{F}} = \exp \left( \hat{u}^\dagger\mathcal{F}^\dagger\hat{v}^\dagger\right)\ket{0} \,,
\label{eq:unitary_vacuum}
\end{equation}
such that
\begin{equation}
\hat{a} \ket{\mathcal{F}} = 0 \qquad \hat{b}  \ket{\mathcal{F}} = 0 \,.
\end{equation}
Acting on it with the new creation operators, it is possible to define a new basis of coherent states as in~\eqref{eq:A_coherent_state_2}:
\begin{equation}
\begin{aligned}
\ket{\alpha\beta;\mathcal{F}} &= \exp\left(-\alpha \hat{a}^\dagger - \beta \hat{b}^\dagger \right) \ket{\mathcal{F}}\\
&= \exp\left(-\alpha \hat{a}^\dagger - \beta \hat{b}^\dagger \right) \exp \left( \hat{u}^\dagger\mathcal{F}^\dagger\hat{v}^\dagger\right)\ket{0} \,,
\end{aligned}
\end{equation}
such that~\footnote{This result corrects equation (2.18) from~\cite{caracciolo:2008ag}}
\begin{equation}
\ket{\alpha\beta;\mathcal{F}} = \exp\left(  \hat{u}^\dagger\mathcal{F}^\dagger\hat{v}^\dagger - \tilde{\alpha}\hat{u}^\dagger - \tilde{\beta} \hat{v}^\dagger - \beta\mathcal{F}\alpha\right)\ket{0}
\label{eq:unitary_coherent_new}
\end{equation}
with
\begin{equation}
\tilde{\alpha} = \mathcal{R}^{-1/2}\alpha \,, \qquad \tilde{\beta} = \beta \mathring{\mathcal{R}}^{-1/2} \,.
\end{equation}
The states constructed acting with the new creation operators on the vacuum $\ket{\mathcal{F}}$ are called, in analogy with condensed matter, \emph{quasiparticle states}. They are eigenvectors for the operators $\hat{a}$, $\hat{b}$:
\begin{equation}
\begin{aligned}
\hat{a} \ket{\alpha\beta;\mathcal{F}} & =  \alpha \ket{\alpha\beta;\mathcal{F}}, \quad &  \hat{b} \ket{\alpha\beta;\mathcal{F}} & =  \beta  \ket{\alpha\beta;\mathcal{F}},\\
\bra{\alpha\beta;\mathcal{F}}\hat{a}^\dagger  & = \bra{\alpha\beta;\mathcal{F}}\alpha^\dagger, \quad & \bra{\alpha\beta;\mathcal{F}}\hat{b}^\dagger  & = \bra{\alpha\beta;\mathcal{F}}\beta^\dagger.
\end{aligned}
\label{eq:unitary_coherent_eigenvalues}
\end{equation}
Further useful properties are collected in Appendix~\ref{subsec:appendixBogoliubov}. 

We can now evaluate the fermion partition function using the basis of the new coherent states:
\begin{equation}
\begin{aligned}
\mathcal{Z}_F &= \Tr^F \prod_t \mathcal{J}_t\hat{\mathcal{T}}_{t,t+1}\\
&= \int \! \fundiff{\alpha^\dagger}\!\fundiff\alpha\! \fundiff{\beta^\dagger}\!\fundiff\beta\, e^{-S_0[U;\mathcal{F}] - S_{QP}[\alpha,\beta,U;\mathcal{F}]} \,,
\end{aligned}
\label{eq:effective_partition}
\end{equation}
where the $S_0$ term accounts for the part not depending on the Grassmannian fields. Using formulas collected in Appendix \ref{subsec:quasi_action}, we can easily write the zero-point action as
\begin{equation}
S_0 [U;\mathcal{F}] = -\sum_t \Tr \left[\ln(\mathcal{R}_t\mathcal{U}_{0,t}\mathcal{E}_{t+1,t})\right]\,,
\label{eq:effective_action_vacuum}
\end{equation}
with the definitions
\begin{subequations}
\begin{gather}
\begin{multlined}[.8\linewidth]
\mathcal{E}_{t+1,t} = \mathcal{F}^\dagger_{\mathcal{N},t+1}e^{\mathcal{M}_{t+1}}\mathcal{U}^\dagger_{0,t}e^{\mathcal{M}^\dagger_t}\mathcal{F}_{\mathcal{N},t} \\
+ \mathcal{F}^\dagger_{t+1}e^{-\mathcal{M}_{t+1}}\mathcal{U}^\dagger_{0,t}e^{-\mathcal{M}^\dagger_t}\mathcal{F}_t \,,
\end{multlined}\\
\begin{multlined}[.8\linewidth]
\mathring{\mathcal{E}}_{t+1,t} = \mathring{\mathcal{F}}_{\mathcal{N},t}e^{\mathcal{M}^\dagger_t}\mathcal{U}_{0,t}e^{\mathcal{M}_{t+1}}\mathring{\mathcal{F}}^\dagger_{\mathcal{N},t+1}\\ 
+ \mathcal{F}_t e^{-\mathcal{M}^\dagger_t}\mathcal{U}_{0,t}e^{-\mathcal{M}_{t+1}}\mathcal{F}^\dagger_{t+1}
\end{multlined}
\end{gather}
\label{eq:effective_E}%
\end{subequations}
and
\begin{equation}
\mathcal{F}_{\mathcal{N},t} = 1 + \mathcal{N}^\dagger_t \mathcal{F}_t\,, \qquad
\mathring{\mathcal{F}}_{\mathcal{N},t} = 1 + \mathcal{F}_t \mathcal{N}^\dagger_t\,.
\label{eq:effective_FN}
\end{equation}
As $S_0$ does not contain the quasiparticle excitations, it can be interpreted as a vacuum contribution. Thus, the parameters of the transformation $\mathcal{F}$, so far generic, can be fixed to the values $\bar{\mathcal{F}}$ that minimize this term, via a variational principle: 
\begin{equation}
\left.\frac{\delta S_0}{\delta \mathcal{F}_t}\right|_{\mathcal{F}_t=\bar{\mathcal{F}}_t} = 0 = \left.\frac{\delta S_0}{\delta \mathcal{F}_t^\dagger}\right|_{\mathcal{F}^\dagger_t=\bar{\mathcal{F}}^\dagger_t} \,.
\label{eq:effective_saddle}
\end{equation}
Of course, because of the presence in $S_0$ of the gauge fields, which evolve in time, those equations have an explicit temporal dependence that makes them difficult to be solved, except for some particular choices of the gauge fields configuration (stationary, or evolving with a pure gauge transformation, see \cite{caracciolo:2010rm}). In the present work we will take a different path, as we will see in section \ref{sec:qcd2_effective}: because of the perturbative nature of confinement in QCD$_2$, where it is simply a consequence of the linearity, in a single spatial dimension, of the Coulomb potential in the distance, a non-trivial result can be obtained even if the gauge fields are eliminated averaging over them after a weak coupling expansion in the coupling constant $g$, truncated at order $g^2$. If this operation is performed \emph{before} arriving at the variational equations \eqref{eq:effective_saddle}, then a corresponding stationary solution can be found, at least numerically, and all the properties of the theory around $\ket{\bar{\mathcal{F}}}$ are valid \emph{on average}, that is after the gauge fields are carefully integrated out.

The quasiparticle action in \eqref{eq:effective_partition} is
\begin{multline}
 S_{QP}[\alpha,\beta,U;\mathcal{F}] = - \sum_t \Bigl[\beta_t \mathcal{I}^{(2,1)}_t \alpha_t + \alpha^\dagger_t \mathcal{I}^{(1,2)}_t\beta^\dagger_t\\
   + \alpha^\dagger_t (\nabla_t - \mathcal{H}_t)\alpha_{t+1} - \beta_{t+1}(\mathring{\nabla}_t -\mathring{\mathcal{H}}_t)\beta^\dagger_t\Bigr]\,,
\label{eq:effective_action_qp}
\end{multline}
where the mixing terms are (if $\det \mathcal{F}\neq 0$)\footnote{These definitions correct the equations (2.31) and (2.32) from~\cite{caracciolo:2008ag}}
\begin{subequations}
\begin{gather}
\begin{multlined}[.8\linewidth]
\mathcal{I}^{(2,1)}_t = \mathring{\mathcal{R}}^{-1/2}_t \Bigl[ \mathring{\mathcal{R}}_t
 - \mathring{\mathcal{E}}^{-1}_{t,t-1}\mathring{\mathcal{F}}_{\mathcal{N},t-1}e^{\mathcal{M}^\dagger_{t-1}}\\
 \cdot\mathcal{U}_{0,t-1}e^{\mathcal{M}_t}\Bigr]\left(\mathcal{F}^\dagger_t \right)^{-1}\mathcal{R}_t^{-1/2} \,,
\end{multlined}\\
\begin{multlined}[.8\linewidth]
\mathcal{I}^{(1,2)}_t =\mathcal{R}^{-1/2}_t \mathcal{F}_t^{-1}\Bigl[\mathring{\mathcal{R}}_t 
- e^{\mathcal{M}^\dagger_t}\mathcal{U}_{0,t}\\
\cdot e^{\mathcal{M}_{t+1}}\mathring{\mathcal{F}}^\dagger_{\mathcal{N},t+1}\mathring{\mathcal{E}}_{t+1,t}^{-1} \Bigr] \mathring{\mathcal{R}}^{-1/2}_t \,,
\end{multlined}
\end{gather}
\label{eq:effective_I}%
\end{subequations}
the one-particle Hamiltonians
\begin{subequations}
\begin{align}
\mathcal{H}_t &= \mathcal{U}_{0,t} - \mathcal{R}_t^{-1/2}\mathcal{E}^{-1}_{t+1,t}\mathcal{R}_{t+1}^{-1/2} \,,\\
\mathring{\mathcal{H}}_t &= \mathcal{U}^\dagger_{0,t} - \mathring{\mathcal{R}}_{t+1}^{-1/2}\mathring{\mathcal{E}}^{-1}_{t+1,t}\mathring{\mathcal{R}}_{t}^{-1/2}
\end{align}
\label{eq:effective_H}%
\end{subequations}
and the lattice derivatives
\begin{subequations}
\begin{align}
\nabla_t &= \mathcal{U}_{0,t} - T^\dagger_0 \,,\\
\mathring{\nabla}_t &= \mathcal{U}^\dagger_{0,t} - T_0 \,.
\end{align}
\label{eq:effective_D}%
\end{subequations}
We will also use
\begin{subequations}
\begin{align}
\mathcal{H}'_t &= 1 - \mathcal{R}_t^{-1/2}\mathcal{E}^{-1}_{t+1,t}\mathcal{R}_{t+1}^{-1/2} \,,\\
\mathring{\mathcal{H}}'_t &= 1 - \mathring{\mathcal{R}}_{t+1}^{-1/2}\mathring{\mathcal{E}}^{-1}_{t+1,t}\mathring{\mathcal{R}}_{t}^{-1/2} \,.
\end{align}
\label{eq:effective_H'}%
\end{subequations}
Evaluated at $\bar{\mathcal{F}}$, the mixing terms are null: the parameters that minimize the vacuum contribution are the ones that decouple positive-energy from negative-energy excitations (see \cite{caracciolo:2008ag}).

\subsection{Mesons}
\label{subsec:mesons}
The previous formulation produces a theory of fermionic excitations, the quasiparticles. However, as long as these excitations are not expected to be present in the low-energy spectrum, an hypothesis of \emph{composite boson dominance} can be formulated: in this phase, all the observable properties of the theory should be understood only in terms of composite bosonic modes, in particular the  lightest among them, the mesons. In our approach, this conjecture can be imposed in a very natural and neat way: the true partition function~\eqref{eq:effective_partition} of the theory should be well approximated by its projection onto states built as condensates of quasiparticles with a mesonic structure. In the following we explain how to perform this operation and how it produces a legitimate action for effective mesons.

\subsubsection{Condensates of quasiparticles}
\label{subsubsec:condensate}

A \emph{quasiparticle condensate} is a state defined by
\begin{equation}
\ket{\Phi;\mathcal{F}} = \exp \bigl(\hat{a}^\dagger \Phi^\dagger \hat{b}^\dagger\bigr) \ket{\mathcal{F}} \,,
\label{eq:mesons_condensate}
\end{equation}
where $\Phi$ is a \emph{structure matrix} with, for now, no further properties. The inner product of such a state with an element of the basis of coherent states is
\begin{equation}
\braket{\alpha\beta;\mathcal{F}| \Phi;\mathcal{F}} = \left(\det \mathcal{R} \right)^{-1} e^{\alpha^\dagger \Phi^\dagger \beta^\dagger}
\label{eq:mesons_inner_coherent}
\end{equation}
so that the norm of the condensate is
\begin{equation}
\braket{\Phi;\mathcal{F}|\Phi;\mathcal{F}} =
\left(\det \mathcal{R} \right)^{-1} \left(\det \mathcal{S} \right)^{-1}\,,
\label{eq:mesons_norm}
\end{equation}
where
\begin{equation}
\mathcal{S} = \left(1+\Phi^\dagger \Phi \right)^{-1} \,, \qquad
\mathring{\mathcal{S}} = \left(1+\Phi\Phi^\dagger \right)^{-1} \,.
\label{eq:mesons_S}
\end{equation}
We would like to evaluate the trace of the transfer matrix only between those states, that is to project on the mesons subspace using the operator
\begin{equation}
\hat{\mathcal{P}}[\mathcal{F}] = \int  \frac{\left[\diff\Phi^\dagger \diff\Phi\right]}{\braket{\Phi;\mathcal{F}|\Phi;\mathcal{F}}} \ket{\Phi;\mathcal{F}}\!\bra{\Phi;\mathcal{F}} \,,
\label{eq:mesons_projector}
\end{equation}
with $\left[\diff\Phi^\dagger \diff\Phi\right]$ a suitable measure in the space of the matrices $\Phi$. Note that this operator is not really a projector: for example,
\begin{equation}
\hat{\mathcal{P}}[\mathcal{F}]^2 \neq \hat{\mathcal{P}}[\mathcal{F}].
\end{equation}
However,
\begin{multline}
\braket{\Phi_t;\mathcal{F}_t| \hat{\mathcal{T}}_{t,t+1}|\Phi_{t+1};\mathcal{F}_{t+1}}\\
\shoveleft{ \quad
= \det \left(e^{-\mathcal{M}^\dagger_t} \mathcal{U}_{0,t} e^{-\mathcal{M}_{t+1}} \mathcal{E}_{t+1,t}\right) 
}
\\
\shoveleft{\qquad \cdot
\det\Bigl\{\left(\mathring{\mathcal{R}}^{-1/2}_{t+1}\mathring{\mathcal{E}}_{t+1,t}^{-1}\mathring{\mathcal{R}}^{-1/2}_{t} \right) \Bigl[ \left(1 + \Phi_t \mathcal{I}^{(1,2)}_t  \right)}\\
\shoveright{{}\cdot\left( \mathring{\mathcal{R}}^{1/2}_{t}\mathring{\mathcal{E}}_{t+1,t}\mathring{\mathcal{R}}^{1/2}_{t+1}\right)\left(1 + \mathcal{I}^{(2,1)}_{t+1}\Phi^\dagger_{t+1}\right)\,}
\\ 
+  \Phi_t
\left( \mathcal{R}^{-1/2}_{t}\mathcal{E}_{t+1,t}^{-1} \mathcal{R}^{-1/2}_{t+1}\right)\Phi^\dagger_{t+1} \Bigr]\Bigr\} \,.
\end{multline}
The partition function restricted on the composites subspace is:
\begin{equation}
\begin{aligned}
\mathcal{Z}_C &= \Tr^F \prod_t \mathcal{J}_t\hat{\mathcal{P}}_t \hat{\mathcal{T}}_{t,t+1}\\
&=  \int \!\prod_t\left[\diff{\Phi^\dagger_t} \diff{\Phi_t}\right] \mathcal{J}_t\frac{\braket{\Phi_t;\mathcal{F}_t | \hat{\mathcal{T}}_{t,t+1} | \Phi_{t+1} ;\mathcal{F}_{t+1}}}{\braket{\Phi_t;\mathcal{F}_t |\Phi_t;\mathcal{F}_t}}\\
&= \int \! \fundiff{\Phi^\dagger}\!\fundiff\Phi \, e^{-S_0[U;\mathcal{F}] - S_M[\Phi,\Phi^\dagger,U;\mathcal{F}]} \,,
\end{aligned}
\label{eq:mesons_partition}
\end{equation}
where the \emph{composite effective action} is
\begin{equation}
S_M[\Phi,\Phi^\dagger,U;\mathcal{F}] = \sum_t \Tr \left\{\ln\left(1 +\Phi^\dagger_t\Phi_t \right) - \ln \mathcal{D}_{t,t+1} \right\} \,,
\label{eq:mesons_action}
\end{equation}
with
\begin{multline}
\mathcal{D}_{t,t+1} =  1 + \mathcal{I}^{(2,1)}_{t+1}\Phi^\dagger_{t+1}\\
+ \left(\mathring{\mathcal{R}}^{-1/2}_{t+1}\mathring{\mathcal{E}}_{t+1,t}^{-1}\mathring{\mathcal{R}}^{-1/2}_{t} \right)\Bigl\{ \Phi_t \mathcal{I}^{(1,2)}_t  \left( \mathring{\mathcal{R}}^{1/2}_{t}\mathring{\mathcal{E}}_{t+1,t}\mathring{\mathcal{R}}^{1/2}_{t+1}\right) \\
\shoveright{+\Phi_t \Bigl[\mathcal{I}^{(1,2)}_t  \left( \mathring{\mathcal{R}}^{1/2}_{t}\mathring{\mathcal{E}}_{t+1,t}\mathring{\mathcal{R}}^{1/2}_{t+1}\right)  \mathcal{I}^{(2,1)}_{t+1} \qquad}\\
+ \left( \mathcal{R}^{-1/2}_{t}\mathcal{E}_{t+1,t}^{-1} \mathcal{R}^{-1/2}_{t+1}\right)
\Bigr]\Phi^\dagger_{t+1}\Bigr\} \,.
\end{multline}

\subsubsection{Colourless pairs in the large \texorpdfstring{$N_c$}{N\_c} limit}
\label{subsubsec:mesonlargeN}
Can the state~\eqref{eq:mesons_condensate} represent a meson? Suppose we can specialize the multi-index $J$ carried by the fermion operators to a set $J=\set{\vec{x},\alpha,i}$ designating, respectively, position, internal (like spin) and colour indices. Then we can define the annihilators and creators for fermion-antifermion colourless composites as the bi-local operators
\begin{equation}
\begin{aligned}
\hat{\Gamma}_{\alpha\beta}(\vec{x},\vec{y}) &= \frac{1}{\sqrt{N_c}}\sum_{i=1}^{N_c} \hat{b}_\alpha^i(\vec{x}) \hat{a}_\beta^i(\vec{y}) , \\ 
\hat{\Gamma}^\dagger_{\alpha\beta}(\vec{x},\vec{y}) &= \frac{1}{\sqrt{N_c}}\sum_{i=1}^{N_c}\hat{a}^{\dagger i}_\alpha(\vec{x}) \hat{b}^{\dagger i}_\beta(\vec{y}).
\end{aligned}
\end{equation}
Note that, with our notations, $\hat{\Gamma}^\dagger_{\alpha\beta}(\vec{x},\vec{y}) \neq \left[\hat{\Gamma}_{\alpha\beta}(\vec{x},\vec{y})\right]^\dagger$ (other possible conventions in~\cite{bardeen:1988mh,kalashnikova:2001df}).
These operators  are not properly canonical operators for bosons, \emph{unless $N_c$ is infinite}, because they are nihilpotent, that is
\begin{equation}
\left[\hat{\Gamma}^\dagger_{\alpha\beta}(\vec{x},\vec{y})\right]^{N_c +1} = 0
\label{eq:mesons_nilpotency}
\end{equation}
(as $\hat{a}^\dagger$, $\hat{b}^\dagger$ are in the fundamental representation of SU($N_c$), there are only $N_c$ components available, and $(\hat{a}_i)^2= 0 = (\hat{b}_i)^2$), and they do not satisfy canonical commutation relations, but 
\begin{multline}
\left[\hat{\Gamma}_{\alpha\beta}(\vec{x},\vec{y}), \hat{\Gamma}^\dagger_{\gamma\delta}(\vec{z},\vec{u}) \right]
= \delta_{\alpha\delta}\delta_{\gamma\beta}\delta(\vec{x}-\vec{u})\delta(\vec{z}-\vec{y}) \\
- \frac{1}{N_c}\sum_i \Bigl[\delta_{\alpha\delta}\delta(\vec{x}-\vec{u})\hat{a}^{\dagger i}_{\gamma}(\vec{z}) \hat{a}^{i}_\beta(\vec{y}) \\
 +  \delta_{\gamma\beta}\delta(\vec{z}-\vec{y})\hat{b}^{\dagger i}_\delta(\vec{u})\hat{b}^{i}_\alpha(\vec{x})\Bigr]
\end{multline}
which become canonical only when $N_c \to \infty$.

Is $\ket{\Phi;\mathcal{F}}$ an eigenstate of $\hat{\Gamma}$ in the same way as~\eqref{eq:A_coherent_eigenstate} and~\eqref{eq:unitary_coherent_eigenvalues} hold? In other terms, is $\ket{\Phi;\mathcal{F}}$ a coherent bosonic state for the operator $\hat{\Gamma}$? It holds that (sum over repeated indices is understood)
\begin{multline}
\hat{\Gamma}_{\alpha\beta}(\vec{x},\vec{y}) \ket{\Phi;\mathcal{F}} = \frac{1 }{\sqrt{N_c}} \Bigl[  \Phi^{\dagger ii}_{\beta \alpha}(\vec{y},\vec{x}) \\
+ \Phi^{\dagger ik}_{\beta\delta}(\vec{y},\vec{u})\hat{b}^{\dagger k}_\delta(\vec{u}) \hat{a}^{\dagger j}_\gamma(\vec{z}) \Phi^{\dagger ji}_{\gamma\alpha}(\vec{z},\vec{x})\Bigr] \ket{\Phi;\mathcal{F}}.
\end{multline}
Suppose now that $\Phi$ has the colour structure
\begin{equation}
\Phi = \mathbb{I}_{N_c} \frac{\phi}{\sqrt{N_c}} \,,
\end{equation}
with $\phi$ a matrix in space and spin indices which carries no gauge index. Then
\begin{multline}
\hat{\Gamma}_{\alpha\beta}(\vec{x},\vec{y}) \ket{\Phi;\mathcal{F}} = \phi^\dagger_{\beta\alpha}(\vec{y},\vec{x}) \ket{\Phi;\mathcal{F}} \\
- \frac{1}{N_c} \phi^\dagger_{\beta\delta}(\vec{y},\vec{u}) \hat{\Gamma}^\dagger_{\delta\gamma}(\vec{u},\vec{z}) \phi^\dagger_{\gamma\alpha}(\vec{z},\vec{x}) \ket{\Phi;\mathcal{F}}.
\end{multline}
The second term is $O(1/N_c)$ and so, when $N_c$ is large,
\begin{equation}
\hat{\Gamma} \ket{\Phi;\mathcal{F}} \simeq \phi^* \ket{\Phi;\mathcal{F}},
\end{equation}
that is, $\ket{\Phi;\mathcal{F}}$ becomes a canonical bosonic coherent state with (matrix) eigenvalue $\phi^* $. Indeed, it can be written as
\begin{equation}
\ket{\Phi;\mathcal{F}} = \exp\left[\phi^*_{\alpha\beta}(\vec{x},\vec{y}) \hat{\Gamma}^\dagger_{\beta\alpha}(\vec{y},\vec{x}) \right] \ket{\mathcal{F}} \,.
\end{equation}
In this limit it is also true that
\begin{equation}
\hat{\mathcal{P}}[\mathcal{F}]\ket{\Phi';\mathcal{F}} =
 \ket{\Phi';\mathcal{F}}\,,
 \label{eq:mesons_projector_largeN}
\end{equation}
so $\hat{\mathcal{P}}[\mathcal{F}]$ becomes a true projector in the mesons subspace. The meson effective action \eqref{eq:mesons_action}, when $N_c$ is large, takes the simple polynomial form
\begin{multline}
S_M = \frac{1}{N_c}\sum_t \Tr \biggl[ \Bigl( \phi^\dagger_t \phi_t - \phi_t \phi^\dagger_{t+1}\Bigr) \\
 +\Bigl(\mathring{\mathcal{H}}'_t \phi_t \phi^\dagger_{t+1} + \mathcal{H}'_t \phi^\dagger_{t+1}\phi_t  \Bigr)  + \frac{1}{2}\Bigl(-2 \phi^\dagger_{t+1} \mathring{\mathcal{H}}'_t \phi_t \mathcal{H}'_t \\
\shoveright{ + \phi_t \mathcal{I}^{(1,2)}_t \phi_t \mathcal{I}^{(1,2)}_t  + \phi^\dagger_{t}\mathcal{I}^{(2,1)}_{t}\phi^\dagger_{t}\mathcal{I}^{(2,1)}_{t} \Bigr)\biggr]}\\
 - \frac{1}{\sqrt{N_c}}\sum_t \Tr \left(\phi_t \mathcal{I}^{(1,2)}_t + \phi^\dagger_{t}\mathcal{I}^{(2,1)}_{t} \right) \,.
\label{eq:mesons_action_thooft}
\end{multline}
As already said before, when the variational principle fixes the parameter $\mathcal{F}$ of the Bogoliubov transformation in order to minimize the vacuum energy contribution, the mixing terms defined in~\eqref{eq:effective_I} are null, and so is the last line of~\eqref{eq:mesons_action_thooft}, leaving a quadratic action for the mesonic bilocal fields $\phi$. This is true also in QCD$_2$, after the average over the gauge configurations is performed. We will explain this procedure in section \ref{sec:qcd2_effective}.

\subsubsection{Relationship with an expansion around a saddle point}
In this subsection we will clarify the connection between the bosonization method explained above and the similar previous point of view presented for example in~\cite{caracciolo:2006wc}. 
In both approaches there is a Bogoliubov transformation of the same form as~\eqref{eq:unitary_bogoliubov} with  $\mathcal{F}$ fixed by the variational principle~\eqref{eq:effective_saddle}.
Within this paper, the  states~\eqref{eq:mesons_condensate} are constructed by means of a second  Bogoliubov transformation of the same form as~\eqref{eq:unitary_bogoliubov}, parametrized by  a fluctuating $\Phi$ which, following the reasoning in paragraph~\ref{subsubsec:mesonlargeN}, is suppressed with respect to $\mathcal{F}$ in the limit of large $N_c$ (which, in the present scheme, has the role of a nilpotency index, see~\eqref{eq:mesons_nilpotency}):
\begin{equation}
\ket{0} \quad\overset{\mathcal{F}}{\longrightarrow}\quad \ket{\mathcal{F}} \quad\overset{\Phi}{\longrightarrow}\quad \ket{\Phi;\mathcal{F}}\,,
\label{eq:composition_vacuum}
\end{equation}
and these states can be interpreted as bosonic excitations of the vacuum made of quasiparticles.

In~\cite{caracciolo:2006wc},  the bosonic excitations are obtained through a \emph{single} Bogoliubov transformation, written as
\begin{equation}
\ket{0} \quad \xrightarrow{\mathcal{F} + \delta\mathcal{F}}\quad \ket{\mathcal{F} + \delta \mathcal{F}} \,,
\label{eq:composition_vacuum_single}
\end{equation}
where now $\delta\mathcal{F}$ are oscillations suppressed for large nilpotency index. This translates into evaluating the resulting fermionic path integral to second order in $\delta\mathcal{F}$ around the saddle point $\mathcal{F} = \bar{\mathcal{F}}$.

Are the two descriptions equivalent? To answer this, we note that, in the basis~\eqref{eq:transfer_std_basis}, a Bogoliubov transformation of the form~\eqref{eq:unitary_bogoliubov} can be written in general as a unitary linear transformation on the vector space of Dirac operators:
\begin{equation}
\hat{\psi}' =  U\, \hat{\psi} \, , \qquad U = V R^{1/2} (1 + F) \,,
\label{eq:composition_Bogo}
\end{equation}
where
\begin{equation}
F = \begin{pmatrix}
0 & -\mathcal{F}^\dagger \\
\mathcal{F} & 0
\end{pmatrix}, \quad R = \begin{pmatrix}
\mathcal{R} & 0 \\
0 & \mathring{\mathcal{R}}
\end{pmatrix}
\end{equation}
and $V$ is a unitary matrix which does not mix creation and annihilation operators:
\begin{equation}
V = \begin{pmatrix}
\mathcal{V}^+	& 0 \\
0				& \mathcal{V}^-
\end{pmatrix}\, ,
\end{equation}
with $\mathcal{V}^+$, $\mathcal{V}^-$ unitary matrices. More generally, given $\hat{\psi}$ in any basis and given an involution $I$,
\begin{equation}
I^\dagger = I\,, \qquad I^2 = \mathbb{I}\,,
\end{equation}
which defines the components ``$+$'' and ``$-$'' of $\hat{\psi}$ via the corresponding projection operators,
\begin{equation}
P_{I}^\pm = \frac{\mathbb{I} \pm I}{2} \quad\implies\quad \hat{\psi}^\pm \equiv P_{I}^\pm \hat{\psi}\, ,
\end{equation}
so that the vector space $W$ where the fermion operator $\hat{\psi}$ leaves can be decomposed as $W^{I,+} \oplus W^{I,-}$, a unitary transformation $U$ on $W$ can always be decomposed as
\begin{equation}
U = U_{e} + U_{o}\, ,
\end{equation}
where $U_{e}$ ($U_{o}$) commutes (anti-commutes) with $I$:
\begin{equation}
\begin{aligned}
I U_{e} I &= U_{e}\\
I U_{o} I &= - U_{o}
\end{aligned} \quad\implies\quad 
\begin{aligned}
U_{e} &: W^{I,\pm} \to W^{I,\pm} \\
U_{o} &: W^{I,\pm} \to W^{I,\mp} \,. 
\end{aligned}
\end{equation}
In the basis where $I = \text{diag}\{1,-1\}$,
\begin{equation}
U_{e} = \begin{pmatrix}
U_+		&	0\\
0		&	U_-
\end{pmatrix}, \qquad U_{o} = \begin{pmatrix}
0		&	U_{+-}\\
U_{-+}		&	0
\end{pmatrix}.
\end{equation}
The choice~\eqref{eq:composition_Bogo}, which we always make in the present work, corresponds to the choice of the involution
\begin{equation}
I = \gamma^0\,.
\end{equation}
The unitary transformations~\eqref{eq:composition_Bogo} form a group:
\begin{equation}
\hat{\psi}'' = U_{2}\, U_{1}\, \hat{\psi} = U_{(2,1)}\, \hat{\psi} \, ,
\end{equation}
where we denote with $(\cdot\, ,\cdot)$ the composition rule. The above composition \eqref{eq:composition_vacuum} is obtained with $\mathcal{F}_1 = \mathcal{F}$, $\mathcal{F}_2 = \Phi$. Note that it is always possible to choose $\mathcal{F}_1$, $\mathcal{F}_2$ in order to keep the block-diagonal terms $V_1$ and $V_2$ as the identity matrix, as we did above, but then $V_{(2,1)}$ is fixed. In this case, the composition rule is
\begin{equation}
\begin{gathered}
F_{(\Phi,\mathcal{F})} = \begin{pmatrix}
0 & -\mathcal{G}^\dagger \\
\mathcal{G} & 0
\end{pmatrix}, \quad R_{(\Phi,\mathcal{F})} = \begin{pmatrix}
\mathcal{Q} & 0 \\
0 & \mathring{\mathcal{Q}}
\end{pmatrix} , \\
V_{(\Phi,\mathcal{F})} = \begin{pmatrix}
\mathcal{A}^+	& 0\\
0				& \mathcal{A}^-
\end{pmatrix}\, ,
\end{gathered}
\end{equation}
with
\begin{equation}
\begin{aligned}
& \mathcal{G} = \mathcal{F}  +   \mathring{\mathcal{R}}^{-1/2}   \left( 1  - \Phi\mathcal{F}^\dagger \right)^{-1}\Phi \mathcal{R}^{-1/2} \, ,\\
& \mathcal{Q} = \left(1+\mathcal{G}^\dagger\mathcal{G} \right)^{-1} = \mathcal{R}^{1/2}   \left( 1  - \mathcal{F}^\dagger \Phi \right) \mathcal{S} \left( 1  - \Phi^\dagger\mathcal{F} \right) \mathcal{R}^{1/2}\, , \\
& \mathring{\mathcal{Q}} = \left(1+\mathcal{G}\mathcal{G}^\dagger \right)^{-1} = \mathring{\mathcal{R}}^{1/2}   \left( 1  - \mathcal{F}\Phi^\dagger \right) \mathring{\mathcal{S}} \left( 1  - \Phi\mathcal{F}^\dagger \right) \mathring{\mathcal{R}}^{1/2} \, , \\
& \mathcal{A}^+ = \mathcal{S}^{1/2} \left(1 - \Phi^\dagger \mathcal{F} \right) \mathcal{R}^{1/2} \, ,\\
& \mathcal{A}^- = \mathring{\mathcal{S}}^{1/2} \left( 1  - \Phi \mathcal{F}^\dagger \right) \mathring{\mathcal{R}}^{1/2} \, .
\end{aligned}
\end{equation}
Given that, we can conclude that, factor a block-diagonal transformation, the composed transformation can be written as a single one with parameters $\mathcal{F} + \delta \mathcal{F}$, such that
\begin{equation}
\begin{aligned}
\delta \mathcal{F} &=  \mathring{\mathcal{R}}^{-1/2}   \left( 1  - \Phi\mathcal{F}^\dagger \right)^{-1}\Phi \mathcal{R}^{-1/2} \\
&= \mathring{\mathcal{R}}^{-1/2} \Phi \mathcal{R}^{-1/2} + O(\Phi^2)\,.
\end{aligned}
\label{eq:composition_deltaF}
\end{equation}
This also means that, when opportunely normalized, the states $\ket{\mathcal{F} + \delta \mathcal{F}}$ and $\ket{\Phi;\mathcal{F}}$ differ only for a phase factor, and so they represent the same physical state:
\begin{equation}
\frac{c}{|c|}\, \frac{ \ket{\mathcal{F} + \delta \mathcal{F}}}{\sqrt{\braket{\mathcal{F} + \delta \mathcal{F}|\mathcal{F} + \delta \mathcal{F}}}} =  \frac{\ket{\Phi;\mathcal{F}}}{\sqrt{\braket{\Phi;\mathcal{F}|\Phi;\mathcal{F}}}} \, ,
\end{equation}
with
\begin{equation}
c = \exp\left[-\Tr\ln\left( 1  +\mathring{\mathcal{R}} \mathcal{F}\delta\mathcal{F}^\dagger  \right) \right] \, .
\end{equation}
In this way we can see that the two procedures give an equivalent description of the resulting mesonic theory. Note that this is not a trivial observation, because in~\eqref{eq:composition_vacuum} the mesonic condensate is built using the quasiparticle operators~\eqref{eq:unitary_bogoliubov}, while in~\eqref{eq:composition_vacuum_single} only the fundamental $\hat{u}^\dagger$, $\hat{v}^\dagger$ appear. However, from the point of view of physical interpretation, we find more natural to describe the excitations above the non-perturbative vacuum $\ket{\mathcal{F}}$ using the corresponding quasiparticles, so that the approach with two distinct Bogoliubov transformations follows. Moreover, the explicit construction of the projector~\eqref{eq:mesons_projector} and the request~\eqref{eq:mesons_projector_largeN} fix the measure over the structure matrices $\left[\diff \Phi^\dagger \diff \Phi \right]$ to be flat in the matrix elements, whereas, in order to set up the saddle point expansion in~\cite{caracciolo:2006wc}, an arbitrary probability measure $\diff \mu(\mathcal{F},\mathcal{F}^\dagger)$ can be used to define the path integral in $\mathcal{Z}_C$ (see~\cite{caracciolo:2010rm,caracciolo:2010xe,caracciolo:2011aa}). 
We leave a deeper understanding of the relation between the two approaches to future investigations.

\section{\label{sec:qcd2}\texorpdfstring{QCD$_{2}$}{QCD2} on the lattice}

In this section we present the 't~Hooft model~\cite{tHooft:1974pnl}, in order to test how far it is possible to follow practically the method explained previously. In doing so, we were inspired by the early work~\cite{bardeen:1988mh}, which adopts a point of view quite consistent with ours, but in the setting of continuum QFT in Hamiltonian approach. For a full treatment of this model along those lines, see~\cite{kalashnikova:2001df}.

\subsection{Wilson action}

The action for the model consists of two distinct terms: a fermion and a pure gauge ones, such that
\begin{equation}
S = S_F + S_G \,.
\end{equation}
The fermion part of the action is
\begin{multline}
S_F = a_0 a_1 \sum_{x\in (a_0\mathbb{Z})\times(a_1\mathbb{Z})}\Biggl\{\left(m+\frac{r_0}{a_0}+\frac{r_1}{a_1}\right)\bar{\psi}(x)\psi(x) \\
-\sum_{\mu=0}^1\biggl[\bar{\psi}(x)\frac{r_\mu-\gamma_\mu}{2a_\mu}U_\mu (x)\psi(x+a_\mu\hat{\mu}) \\
+ \bar{\psi}(x+a_\mu\hat{\mu})\frac{r_\mu+\gamma_\mu}{2a_\mu}U^\dagger_\mu (x) \psi(x)\biggr]
\Biggr\} \,.
\label{eq:QCD2_action_fermion_no_ais}
\end{multline}
The terms proportional to the Wilson parameters $r_\mu$ are introduced to solve the fermion doubling problem. The gauge link operators are
\begin{equation}
U_0(x) = e^{iga_0 A_0(x) }\,, \qquad U_1(x) = e^{iga_1 A_1 (x)}\,,
\label{eq:QCD2_gauge_U}
\end{equation}
with the algebra-valued Hermitian field~\footnote{We use the ``early'' latin letters $a$, $b$, $\cdots$ to indicate the $N^2_c -1$ indices of the adjoint representation of the gauge group, while we assign the ``late'' letters $j$, $k$, $\cdots$ to the $N_c$ indices of the fundamental one. However, as long as possible, we omit the latter to lighten notation: when not explicitly written, they are contracted following the natural order. ``$\tr$'' is a trace over the fundamental indices.}
\begin{equation}
A_\mu (x) = \sum_{a=1}^{N_c^2 - 1} A_\mu^a(x) \Theta^a
\label{eq:QCD2_gauge_A}
\end{equation}
and $\Theta^a$ a set of Hermitian generators of the algebra, normalized such that
\begin{equation}
\tr\left(\Theta^a \Theta^b \right) = \frac{1}{2}\delta_{ab} \,.
\end{equation}
The quadratic Casimir operator is
\begin{equation}
\Theta\cdot\Theta \equiv \sum_{a=1}^{N^2_c - 1} \Theta^a \Theta^a = \frac{N_c^2 - 1}{2N_c}\, \mathbb{I}_{N_c} \,.
\label{eq:QCD2_Casimir}
\end{equation}
We will use also the following Fierz-type identity (see, for example,~\cite{nishi:2004st}):
\begin{equation}
\sum_{a=1}^{N^2_c - 1} \Theta^a_{ij} \Theta^a_{kl} = \frac{1}{2}\left(\delta_{il}\delta_{jk} - \frac{1}{N_c} \delta_{ik}\delta_{jl} \right) \,.
\label{eq:QCD2_Fierz}
\end{equation}
We call the representation of the action \eqref{eq:QCD2_action_fermion_no_ais} ``mass form'', because the bare mass parameter $m$ is explicit in it. Introducing the \emph{hopping parameters}
\begin{equation}
\begin{aligned}
\kappa_0 &\equiv \frac{1}{2}\left(m a_0  + r_0 + r_1\frac{a_0}{a_1}\right)^{-1} \,, \\ 
\kappa_1 &\equiv \frac{1}{2}\left(m a_1  + r_0\frac{a_1}{a_0} + r_1\right)^{-1} \,,
\end{aligned}
\end{equation}
so that
\begin{equation}
\zeta \equiv \frac{a_0}{a_1} = \frac{\kappa_1}{\kappa_0} \,,
\end{equation}
and rescaling the fermion fields into the dimensionless variables
\begin{equation}
\varphi = \left(\frac{a_1}{2\kappa_0} \right)^{1/2} \psi\,, \qquad \bar{\varphi} = \left(\frac{a_1}{2\kappa_0} \right)^{1/2} \bar{\psi} \,,
\label{eq:QCD2_scaling_hopping}
\end{equation}
the action can be written in the ``hopping-parameter form''
\begin{multline}
S_F = \sum_{n\in\mathbb{Z}}\sum_{t\in \mathbb{Z}}\Biggl\{\bar{\varphi}_{n,t}\varphi_{n,t} \\
\shoveleft{\quad - \kappa_1\biggl[\bar{\varphi}_{n,t} (r_1 - \gamma^1) (U_1)_{n,t}\varphi_{n+1,t} }\\[-1.2ex]
\shoveright{+ \bar{\varphi}_{n+1,t}  (r_1 + \gamma^1) (U^\dagger_1)_{n,t} \varphi_{n,t}\biggr]\phantom{\Biggr\}}}\\[-1ex]
\shoveleft{\quad- \kappa_0\biggl[\bar{\varphi}_{n,t}(r_0-\gamma^0)(U_0)_{n,t}\varphi_{n,t+1} }\\[-1.2ex]
+ \bar{\varphi}_{n,t+1}(r_0+\gamma^0)(U^\dagger_0)_{n,t} \varphi_{n,t}\biggr]
\Biggr\} \,,
\label{eq:QCD2_action_fermion_hopping}
\end{multline}
with dimensionless lattice units
\begin{equation}
t = x_0/a_0 \,, \qquad n = x_1/a_1 \,.
\end{equation}
To avoid inessential complications due to time doublers in the construction of the fermionic transfer matrix, which we will discuss below, we will always take $r_0=1$, so that the operators $r_0\pm\gamma_0$ become projectors.~\footnote{For a construction of the fermionic transfer matrix with a generic $r_0$, using a double time slice Hilbert space formalism, see~\cite{smit:1990pb}.}

\subsection{Coulomb gauge}
\label{subsec:coulomb}
A standard choice of the pure gauge action, that reduces to the Yang-Mills action in the continuum limit, is
\begin{equation}
S_G = \frac{1}{a_0a_1}\sum_P\frac{1}{g^2}\left[2N_c - \Tr\left(U_P + U_P^\dagger\right)\right] \,,
\label{eq:QCD2_action_gauge}
\end{equation}
where the plaquette sum and variables are defined by
\begin{equation}
\begin{aligned}
\sum_P &= \frac{1}{2}\sum_{x} \sum_{\mu,\nu=0}^1\,,\\
U_P &= U_\mu(x)U_\nu(x+a_\mu\hat{\mu})U^\dagger_\mu (x+a_\nu\hat{\nu})U^\dagger_\nu(x) \,.
\end{aligned}
\label{eq:QCD2_gauge_plaquette}
\end{equation}
As it is clear from~\eqref{eq:QCD2_action_gauge}, the bare coupling constant $g$ in $\text{QCD}_2$ has dimension of mass (the theory is super-renormalizable) while the field $A_\mu$ is dimensionless.

The gauge action can be written as a quadratic plus an interaction part in the field $A_\mu(x)$ only in the continuum limit, when it becomes the usual Yang-Mills action for gluons. In order to define a gluon propagator for any finite lattice spacing $a$ we need to formulate a perturbation theory in $g$ on the lattice, the so called \emph{weak coupling expansion}, mirroring the diagrammatic series in the continuum (see the details in textbooks, like~\cite{montvay:lqft,rothe:lgt}).
For our future convenience, we only need an expression for the gluon propagator at the lowest order in $g$ (that is, at tree level) in the Coulomb gauge. This gauge in two space-time dimensions is fixed by the condition
\begin{equation}
U_1(x) = 1 \quad\iff\quad A_1(x) = 0 \,.
\label{eq:weak_Coulomb_gauge}
\end{equation}
so that the only gluon field remaining in the theory is $A_0$.
Thus, the free gluon propagator is
\begin{equation}
\mathcal{G}^{ab}_{00}(x,y) \equiv \braket{A_0^a(x) A^b_0(y)} = - \frac{1}{a_0 a_1}\delta_{ab} \left[(\partial_1)^{-2}\right]_{xy} .
\end{equation}
The brackets indicate the average over the gauge fields. Using the Fourier transform~\eqref{eq:A_Fourier} we get the momentum representation
\begin{equation}
\begin{aligned}
\mathcal{G}^{ab}_{00}(p,q)&= (2\pi)^2 \delta^{(2)}(p+q)\mathcal{G}^{ab}_{00}(p) \,,\\
\mathcal{G}^{ab}_{00}(p) &\equiv \delta_{ab}\frac{1}{(\hat{p}^1)^2} \,,
\end{aligned}
\label{eq:weak_propagator_momentum}
\end{equation}
with
\begin{equation}
\hat{p}^1 = \frac{2}{a_1}\sin \frac{a_1 p^1 }{2} \,,
\label{eq:weak_momentum_lattice}
\end{equation}
so that
\begin{equation}
\mathcal{G}^{ab}_{00}(x,y) = \delta_{ab}\frac{\delta_{x^0 , y^0}}{a_0}\int_{-\pi/a_1}^{\pi/a_1} \frac{\diff p}{2\pi}\,\frac{e^{ip\left(x^1-y^1\right)}}{\hat{p}^2} \,.
\label{eq:weak_propagator}
\end{equation}
In the continuum limit $a_0, a_1 \to 0$ this equation gives the well known result
\begin{equation}
\begin{aligned}
\mathcal{G}^{ab}_{00}(x,y) &= \delta_{ab}\delta(x^0 - y^0)\int_{-\infty}^{+\infty} \frac{\diff p}{2\pi}\,\frac{e^{ip\left(x^1-y^1\right)}}{p^2}\\
&= -\frac{1}{2}\delta_{ab}\delta(x^0 - y^0)\left|x^1 - y^1 \right|\,,
\label{eq:weak_propagator_continuum}
\end{aligned}
\end{equation}
where the last equality can be obtained with a prescription to regularize the integral (see, for reference,~\cite{kalashnikova:2001df}). In this form, it is more evident that, in two space-time dimensions, the interaction between quarks mediated by the gluons yields to linear confinement already in a perturbative setting.

\subsection{Transfer matrix}
\label{subsec:qcd2_transfer}
From the fundamental work~\cite{luescher:1976ms} it is easy to get, for the matrices defining the operator~\eqref{eq:transfer_matrix} in the present model,
\begin{align}
\mathcal{B}_t &\equiv 2\kappa_0 e^{2\mathcal{M}_t} = \mathbb{I} - \kappa_1 r_1 \left(\mathcal{U}_{1,t} T_1 + T^\dagger_1\mathcal{U}^\dagger_{1,t} \right) \,, \label{eq:transfer_B}\\
\mathcal{N}_t &= -i\kappa_1\mathcal{B}_t^{-1/2} \left(\mathcal{U}_{1,t} T_1 - T^\dagger_1\mathcal{U}^\dagger_{1,t}\right)  \mathcal{B}_t^{-1/2}\,, \label{eq:transfer_N}
\end{align}
where we used the definition~\eqref{eq:A_lattice_shift} for the lattice shift operators. In Coulomb gauge~\eqref{eq:weak_Coulomb_gauge} these matrices have the easy, time-independent form 
\begin{align}
\mathcal{B} &= \mathbb{I} - \kappa_1 r_1 (T_1 + T^\dagger_1) \,,\\
\mathcal{N} &= -i\kappa_1\mathcal{B}^{-1/2} \left(T_1 - T^\dagger_1\right)  \mathcal{B}^{-1/2} \,.
\end{align}
Using the Fourier representation~\eqref{eq:A_Fourier} we get
\begin{align}
\mathcal{B}(p,q) &= 2\pi a_1 \delta(p+q) \mathcal{B}(q) \,,\\
\mathcal{N}(p,q) &= 2\pi a_1 \delta(p+q) \mathcal{N}(q) \,, \\
e^{2\mathcal{M}}(p,q) &= 2\pi a_1 \delta(p+q) e^{2\mathcal{M}}(q) \,,
\end{align}
with
\begin{align}
\mathcal{B}(q) &\equiv 1 - 2\kappa_1 r_1 \cos a_1 q \,,\\
\mathcal{N}(q) &\equiv 2\kappa_1 \mathcal{B}^{-1}(q) \sin a_1 q \,, \\
e^{2\mathcal{M}}(q) &= \frac{\mathcal{B}(q)}{2\kappa_0} = 1 + ma_0 - r_1 \zeta (\cos a_1 q - 1) \,.
\end{align}
Thus, we can see that in this case $\mathcal{M}$ and $\mathcal{N}$ commute.

\section{\texorpdfstring{QCD$_2$}{QCD2} effective action}
\label{sec:qcd2_effective}

We have now all the ingredients to apply at the present model the machinery explained above. In this section we derive results for the vacuum contribution, the quasiparticles behaviour and, finally, the mesonic sector of the theory.

\subsection{Vacuum energy}
\label{subsec:vacuum}
In order to proceed, we need to parametrize efficiently the matrix $\mathcal{F}_t$ in~\eqref{eq:unitary_bogoliubov}.
Because all the time dependence comes from the residual gauge field $A_0(t)$, which will be integrated out in an intermediate step, we can choose a transformation corresponding to a stationary $\mathcal{F}_t = \mathcal{F}$. Moreover, in two space-time dimensions there is no residual spin index for the Weyl spinors, and we can think as well the transformation to be an identity in colour space. Because of space translational invariance, its Fourier transform can e written as
\begin{subequations}
\begin{align}
\mathcal{F}(p,q) &= 2\pi a_1 \delta(p+q) \mathcal{F}(q) \,,\\
\mathcal{F}(q) &= \tan \frac{\theta_q}{2} = \mathcal{F}^\dagger(q) \,,
\end{align}
\label{eq:vacuum_F}%
\end{subequations}
where $\theta_q$ is the \emph{Bogoliubov-Valatin angle} usually introduced in literature to parametrize the unitary transformation. So, in QCD$_2$ we can use a single parameter for each value of the momentum. In this way
\begin{align}
\mathcal{R}(p,q) &= 2\pi a_1 \delta(p+q) \mathcal{R}(q) \,,\\
\mathcal{R}(q) &= \left(1 + \tan^2 \frac{\theta_q}{2} \right)^{-1} = \cos^2 \frac{\theta_q}{2} \,,
\end{align}
and
\begin{equation}
\begin{aligned}
\cos \theta_q &= \frac{1-\mathcal{F}^2(q)}{1+\mathcal{F}^2(q)} = 2\mathcal{R}(q) -1 \,,\\
\sin \theta_q &= \frac{2\mathcal{F}(q)}{1+\mathcal{F}^2(q)} \,.
\end{aligned}
\label{eq:vacuum_sincos}
\end{equation}

We will work at the lowest orders in lattice spacings and ultimately take the limit $a_0, a_1 \to 0$, to confront with the previous continuum results~\cite{bardeen:1988mh,kalashnikova:2001df}. For this reason, in the following we will omit the hat~\eqref{eq:weak_momentum_lattice} in lattice momenta and extend the integrals defined in the first Brillouin Zone $[-\pi/a_1,\pi/a_1]$ (as in~\eqref{eq:weak_propagator}) to all $\mathbb{R}$. However, in case of ambiguity, the lattice produces a natural momentum cut-off for divergent integrals. Thanks to Wilson's prescription, we can ignore the contributions coming from the fermion doublers, expanding for $a_1 q \simeq 0$ and then taking the limit $a_1\to 0$. We get, from~\ref{subsec:qcd2_transfer},
\begin{subequations}
\begin{align}
e^{\mathcal{M}}(q) &= 1 + \frac{ma_0}{2} + O(a_0 a_1) \,,\\
\mathcal{N}(q) &= \zeta a_1  q + O(a_0 a_1) = a_0 q + O(a_0 a_1) \,.
\end{align}
\label{eq:vacuum_a0_expansion}%
\end{subequations}
We will also need to expand the link variables~\eqref{eq:QCD2_gauge_U} to second order in $g$, using the results sketched in subsection~\ref{subsec:coulomb}. In this way
\begin{equation}
U_0 = 1 + iga_0A_0 - \frac{1}{2} g^2 a_0^2 A_0^2 + \cdots
\label{eq:vacuum_weak_coupling}
\end{equation}
We remark that, unlike the previous one, this is not actually an expansion in the lattice spacing $a_0$: in particular, once we have mediated over gauge fields, the quantity $a^2_0\braket{A_0^2}$ will be not $O(a_0^2)$, but only $O(a_0)$. Indeed, this object is proportional to the gluon propagator we reported in~\eqref{eq:weak_propagator}: one of the $a_0$ factor simplifies with the one in the denominator and leaves an overall term $O(a_0)$. This will happen to all (even) orders in $g$, because the powers of $a_0$ will always be equal to those of $A_0$. What is really in force here is a weak coupling expansion that, as we will see, is nothing else than the \emph{'t~Hooft limit}
\begin{equation}
N_c \to \infty \qquad \text{with $g^2 N_c $ fixed}.
\label{eq:vacuum_tHooft_limit}
\end{equation}

We now report, as an example of this kind of calculations, all the passages needed to get the continuum limit of the vacuum contribution~\eqref{eq:effective_action_vacuum}; the results in the next sections can be obtained in a similar way. The reader not interested in the details can skip to equation \eqref{eq:vacuum_action}. Expanding~\eqref{eq:effective_E} to second order in $g$ we get
\begin{equation}
\mathcal{E}_{t+1,t} =  \mathcal{E}^{(0)} + g\mathcal{E}^{(1)}_t + g^2 \mathcal{E}^{(2)}_t \,,
\end{equation}
with
\begin{multline}
\mathcal{E}^{(0)} = \left(1+\mathcal{F} \mathcal{N} \right)e^{2\mathcal{M}}\left(1+\mathcal{F} \mathcal{N} \right) + \mathcal{F} e^{-2\mathcal{M}}\mathcal{F} \,,\\
\shoveleft{\mathcal{E}^{(1)}_t = - i\left(1+\mathcal{F} \mathcal{N} \right)e^{\mathcal{M}} a_0A_{0,t}e^{\mathcal{M}}\left(1+\mathcal{N} \mathcal{F} \right) }\\
\shoveright{-i\mathcal{F} e^{-\mathcal{M}}a_0A_{0,t} e^{-\mathcal{M}}\mathcal{F} \,,}\\
\shoveleft{\mathcal{E}^{(2)}_t = - \frac{1}{2}\left(1+\mathcal{F} \mathcal{N} \right)e^{\mathcal{M}} a_0^2 A_{0,t}^2e^{\mathcal{M}}\left(1+\mathcal{N} \mathcal{F} \right) }\\
-\frac{1}{2}\mathcal{F} e^{-\mathcal{M}}a_0^2 A_{0,t}^2 e^{-\mathcal{M}}\mathcal{F} \,,
\end{multline}
so that, in momentum space,
\begin{multline}
\mathcal{E}^{(0)}(p,q) = 2\pi a_1 \delta(p+q)\\
\shoveright{\cdot\bigl[\left(1+\mathcal{F} \mathcal{N} \right)e^{2\mathcal{M}}\left(1+\mathcal{F} \mathcal{N} \right)+ \mathcal{F} e^{-2\mathcal{M}}\mathcal{F}\bigr]\!(q) \,,}\\
\shoveleft{\mathcal{E}^{(1)}_t(p,q) = - ia_1a_0\Bigl\{\left[\left(1+\mathcal{F} \mathcal{N} \right)e^{\mathcal{M}} \right]\!\!(-p)}\\
\cdot A_{0,t}(p+q)\left[e^{\mathcal{M}}\left(1+\mathcal{N} \mathcal{F} \right)\right]\!\!(q)\\ 
\shoveright{+ \left[\mathcal{F} e^{-\mathcal{M}}\right]\!\!(-p)\, A_{0,t}(p+q) \left[ e^{-\mathcal{M}}\mathcal{F}\right]\!\!(q)\Bigr\} \,,}\\
\shoveleft{\mathcal{E}^{(2)}_t(p,q) = - \frac{a_1 a_0^2}{2}\int\frac{\diff k}{2\pi}\Bigl\{ \left[\left(1+\mathcal{F} \mathcal{N} \right)e^{\mathcal{M}}\right]\!\! (-p)}\\
 \cdot A_{0,t}(p+k)A_{0,t}(-k +q) \left[ e^{\mathcal{M}}\left(1+\mathcal{N} \mathcal{F} \right)\right]\!\!(q) \\
 +\left[\mathcal{F} e^{-\mathcal{M}}\right]\!\!(-p)\, A_{0,t}(p+k)A_{0,t}(-k+q)\\
 \cdot \left[ e^{-\mathcal{M}}\mathcal{F}\right]\!\!(q) \Bigr\} \,.
\end{multline}
Using~\eqref{eq:vacuum_a0_expansion} and keeping only first-order terms in $a_0$, we get
\begin{multline}
\mathcal{E}^{(0)}(p,q) = 2\pi a_1 \delta(p+q)\bigl\{1 + \mathcal{F}^2(q)\\
\shoveright{+ ma_0\left[1-\mathcal{F}^2(q)\right] + 2 a_0 q \mathcal{F}(q)
\bigr\}\,,}\\
\shoveleft{\mathcal{E}^{(1)}_t(p,q) = - ia_1a_0A_{0,t}(p+q)\left[1 + \mathcal{F}(-p)\mathcal{F}(q)\right] \,,}\\
\shoveleft{\mathcal{E}^{(2)}_t(p,q) = - \frac{a_1 a_0^2}{2}\int \frac{\diff k}{2\pi}A_{0,t}(p+k)A_{0,t}(-k +q)}\\
\cdot\bigl[1 
+\mathcal{F}(-p)\mathcal{F}(q) \bigr] \,.
\end{multline}
Then, we multiply on left and right by $\mathcal{R}^{1/2}$ and use~\eqref{eq:vacuum_sincos}:
\begin{multline}
\mathcal{R}^{1/2}\mathcal{E}^{(0)}\mathcal{R}^{1/2}(p,q) = 2\pi a_1 \delta(p+q)\\
\shoveright{\cdot\left(1 + ma_0 \cos\theta_q + a_0 q \sin\theta_q\right)\,,}\\
\shoveleft{\mathcal{R}^{1/2}\mathcal{E}^{(1)}_t\mathcal{R}^{1/2}(p,q) = - ia_1a_0A_{0,t}(p+q)\cos \frac{\theta_q - \theta_{-p}}{2} \,,}\\
\shoveleft{\mathcal{R}^{1/2}\mathcal{E}^{(2)}_t\mathcal{R}^{1/2}(p,q) = - \frac{a_1 a_0^2}{2}}\\
\cdot\int \frac{\diff k}{2\pi}A_{0,t}(p+k)A_{0,t}(-k +q)\cos \frac{\theta_q - \theta_{-p}}{2} \,.
\label{eq:effective_RER}
\end{multline}
In evaluating the logarithm of the sum of these quantities to first order in $a_0$ we use the usual formula
\begin{multline}
\ln\left[1+(\mathcal{R}^{1/2}\mathcal{E}_{t+1,t}\mathcal{R}^{1/2} - 1)\right] \simeq \mathcal{R}^{1/2}\mathcal{E}_{t+1,t}\mathcal{R}^{1/2} - 1 \\
- \frac{1}{2}(\mathcal{R}^{1/2}\mathcal{E}_{t+1,t}\mathcal{R}^{1/2} - 1)^2
\end{multline}
and note that the only term that can produce an $O(a_0)$ addend when squared is $\mathcal{R}^{1/2}\mathcal{E}^{(1)}\mathcal{R}^{1/2}$, because of the property of $\braket{A_0^2}$ we already mentioned. As a matter of fact,
\begin{multline}
-\frac{1}{2} \left[\mathcal{R}^{1/2}\mathcal{E}^{(1)}_t\mathcal{R}^{1/2}\right]^2(p,q) = \frac{a_1 a_0^2}{2}\int \frac{\diff k}{2\pi} \\\cdot A_{0,t}(p+k)A_{0,t}(-k+q)\mathcal{R}^{1/2}(-p)\mathcal{R}(k)\mathcal{R}^{1/2}(q)\\
\cdot\bigl[
1
+ \mathcal{F}(-p)\mathcal{F}(k)\mathcal{F}(k)\mathcal{F}(q)\\
+\mathcal{F}(k)\mathcal{F}(q) + \mathcal{F}(-p)\mathcal{F}(k)\bigr]\, .
\end{multline}
Summing the $O(g^2)$ terms we get
\begin{multline}
\mathcal{R}^{1/2}\mathcal{E}^{(2)}_t\mathcal{R}^{1/2}(p,q) -\frac{1}{2} \left[\mathcal{R}^{1/2}\mathcal{E}^{(1)}_t\mathcal{R}^{1/2}\right]^2(p,q) \\= -\frac{a_1 a_0^2}{2}\int\frac{\diff k}{2\pi}A_{0,t}(p+k)A_{0,t}(-k +q)\\
\cdot \sin\frac{\theta_{-p}-\theta_k}{2} \sin\frac{\theta_{q}-\theta_k}{2} \,.
\end{multline}
The logarithm becomes
\begin{equation}
\begin{aligned}
&\left[\ln \mathcal{R}^{1/2}\mathcal{E}_{t+1,t}\mathcal{R}^{1/2}\right](p,q) \simeq \\
&\qquad2\pi a_0 a_1\delta(p+q)\left(m \cos\theta_q +  q \sin\theta_q\right)  \\
&\qquad- i a_0 a_1 g A_{0,t}(p+q) \cos \frac{\theta_q - \theta_{-p}}{2}\\
&\qquad- \frac{a_1 a_0^2 g^2}{2}\int \frac{\diff k}{2\pi}A_{0,t}(p+k)A_{0,t}(-k +q)\\
&\hspace{10em}\cdot\sin\frac{\theta_{-p}-\theta_k}{2} \sin\frac{\theta_{q}-\theta_k}{2} \,.
\end{aligned}
\label{eq:vacuum_log_final_gauge}
\end{equation}

It's now time to integrate over the gauge fields, that is to perform the gauge integral in~\eqref{eq:transfer_partition_transfer}. This means, in weak coupling approximation, that we can substitute the gauge fields with their expectation values. Using~\eqref{eq:weak_propagator_momentum}, ultimately we get
\begin{multline}
\braket{\ln \mathcal{R}^{1/2}\mathcal{E}_{t+1,t}\mathcal{R}^{1/2}}(p,q) \simeq \\2\pi a_0 a_1  \delta(p+q)\biggl( m \cos\theta_q +  q \sin\theta_q \\
- \frac{g^2}{2}\int \frac{\diff k}{2\pi}\frac{\Theta\cdot\Theta}{(q-k)^2}\sin^2 \frac{\theta_q - \theta_k}{2} \biggr)\,,
\label{eq:vacuum_log_final}
\end{multline}
which no more depends on time. We used directly the continuum gluon propagator instead of its lattice version to lighten the notation: that's what we would get in the continuum limit, which is implicit in our choice to maintain only the lowest terms in lattice spacing, as explained at the beginning of this section. We also imposed that $\braket{A_0} = 0$: that's true for all the odd products of the gauge fields. Evaluating the trace over space (see~\eqref{eq:A_Fourier_trace}) and colour indices we find
\begin{multline}
\braket{\Tr \left[\ln \mathcal{R}^{1/2}\mathcal{E}_{t+1,t}\mathcal{R}^{1/2}\right]} = a_0 2\pi \delta(0)\frac{N_c}{2\pi} \\
\shoveleft{\cdot\biggl[\int\!\diff q\left(m  \cos\theta_q +  q \sin\theta_q\right) }\\
- \alpha_s\frac{\left(N_c^2-1  \right)}{2N_c}\int\!\diff q\int \!\diff k\,\frac{1}{(q-k)^2}\sin^2 \frac{\theta_q - \theta_k}{2}  \biggr]\,,
\end{multline}
with
\begin{equation}
\alpha_s = \frac{g^2}{4\pi} \,.
\end{equation}
The factor $2\pi\delta(0)$ is simply the length of the lattice in the spatial direction, as we can see from~\eqref{eq:A_Fourier_delta}:
\begin{equation}
2\pi\delta(0) = a_1 \sum_{x^1} 1 \,.
\end{equation}
Summing over $t$ we can finally derive the expression for the vacuum contribution to the action:
\begin{multline}
S_0[\theta] = \braket{S_0[\theta;A_0]} = - \frac{V N_c}{2\pi}\, \biggl[\int \!\diff q\left(m  \cos\theta_q +  q \sin\theta_q\right) \\
-\frac{ \alpha_s\left(N_c^2-1  \right)}{2 N_c}\int\!\diff q\int\!\diff k\,\frac{1}{(q-k)^2}\sin^2 \frac{\theta_q - \theta_k}{2}  \biggr] \,.
\label{eq:vacuum_action}
\end{multline}
The space-time volume
\begin{equation}
V = a_0 a_1 \sum_{x^0} \sum_{x^1} 1
\end{equation}
accounts for the fact that energy is an extensive property; it can be regularized imposing boundaries on the lattice and then perform a thermodynamic limit procedure. We remark here that, in the weak coupling regime $g\to 0$, this result is not trivial (that is, different from what we would obtain in the free theory) only if $N_c \to \infty$, accordingly to the 't~Hooft's limit~\eqref{eq:vacuum_tHooft_limit}. In this way, the dispersion law for the vacuum energy density is
\begin{equation}
\omega^0_q[\theta] = m  \cos\theta_q +  q \sin\theta_q -\frac{\gamma}{2}\int\!\diff k\,\frac{1}{(q-k)^2}\sin^2 \frac{\theta_q - \theta_k}{2}
\label{eq:vacuum_energy}
\end{equation}
with
\begin{equation}
\gamma \equiv \alpha_s\frac{N_c^2-1  }{N_c} \underset{N_c\to \infty}{\longrightarrow} \alpha_s N_c \,.
\end{equation}
Imposing parity on this expression for any value of $\gamma$, we get
\begin{equation}
\omega^0_q = \omega^0_{-q} \quad\iff\quad \theta_q = -\theta_{-q} \,.
\label{eq:vacuum_parity_FR}
\end{equation}
We also note that~\eqref{eq:vacuum_action} is a $O(N_c)$ contribution in the 't~Hooft's limit.

\begin{figure}[htp]
\input{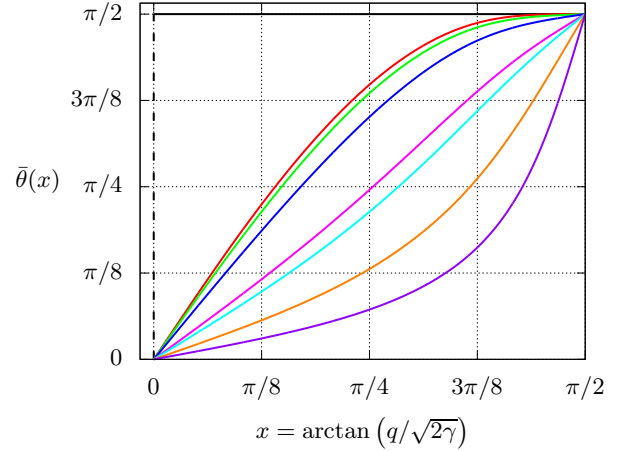}
\caption{Plot for the $\theta_q$ that solves the gap equation. The topmost, discontinuous line (in black) corresponds to the chiral limit of the free theory ($m=0$, $\gamma\to 0$); the following lower one (in red) corresponds to the chiral limit of the interacting theory ($m=0$, $\gamma=1$). The others, from the top to the bottom, correspond to different values of the mass in the interacting theory ($\gamma=1$) in the set $m\in \{0.045,0.18,0.749,1,2.11,4.23\}$, in unities of $\sqrt{2\gamma}$, to confront with references~\cite{li:1987hx,jia:2017uul}.}
\label{fig:theta}
\end{figure}
\begin{figure}[htp]
\input{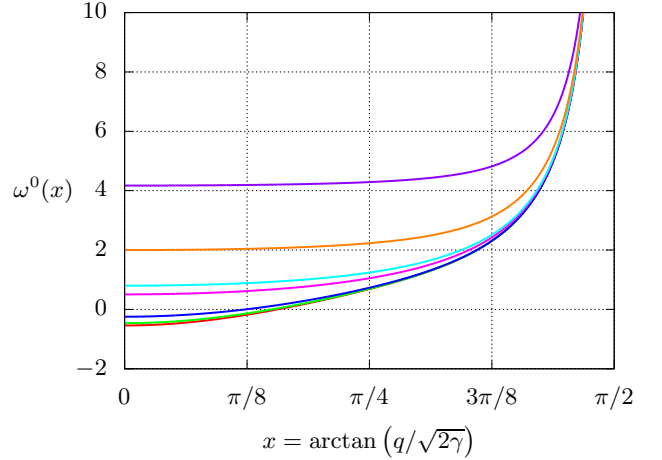}
\caption{Plot for $\omega_q$, evaluated on solutions $\bar{\theta}_q$ of the gap equation. The lines correspond to the mass values used in Figure~\ref{fig:theta}, from the bottom up. }
\label{fig:omega}
\end{figure}

A variation with respect to the Bogoliubov angle $\theta_q$ leads to the gap equation
\begin{equation}
-m  \sin\theta_q +  q \cos\theta_q -\frac{ \gamma}{2}\int\!\diff k\,\frac{\sin \left(\theta_q - \theta_k\right)}{(q-k)^2} = 0 \,.
\label{eq:vacuum_theta_saddle}
\end{equation}
The solution $\bar{\theta}_q$ can be obtained in closed form only in the free ($\gamma=0$) theory, where it is simply
\begin{equation}
\bar{\theta}_q^{\text{(free)}} = \arctan \frac{q}{m} \,.
\label{eq:vacuum_theta_free}
\end{equation}
However, using the numerical methods described in~\cite{jia:2017uul}, we can see its form also in the interacting theory, as reported in Figure~\ref{fig:theta}. Confronting the result with the free value \eqref{eq:vacuum_theta_free}, one can see that the interaction produces, even in the chiral limit, a non-trivial shape for $\bar{\theta}_q$ in a way qualitative similar to the effect of a mass in the free theory. As we will see in a moment, this sort of \emph{dynamical mass} produced by the interaction breaks indeed the chiral symmetry in the massless theory, giving a nonzero value of the chiral condensate (see also \cite{bardeen:1988mh} for a discussion). Once that the values of $\theta_q$ solving the gap equation are available, the form for the energy density $\omega_q^0$, which we interpret as the vacuum contribute to the energy density, follows easily (Figure~\ref{fig:omega}). 

These results are well known in literature: see, for example, reference~\cite{kalashnikova:2001df} and graphs within. The fact that we can derive them in the continuum limit of our present scheme is a proof for the validity of the method proposed.
In particular, equation \eqref{eq:vacuum_theta_saddle} is the celebrated \emph{gap equation for the 't~Hooft model}, written in this form for the first time by Bars and Green in~\cite{bars:1977ud}, both from canonical and diagrammatic approaches. Their result follows from the request for the theory to be diagonal in what we call quasiparticles excitations, that is to decouple quasi-fermions from quasi-antifermions at the quadratic level (when the integration over the gauge fields is performed, the resulting theory is actually quartic in the quasiparticles: see subsection \ref{subsec:quasiparticle}). In our notation, this means to look for the $\bar{\theta}$ that cancels the mixings $\mathcal{I}^{(1,2)}$, $\mathcal{I}^{(2,1)}$ in the quasiparticle action~\eqref{eq:effective_action_qp}. As we noted in section~\ref{sec:bogoliubov} and we will explicitly see in the following, our variational approach produces the requested diagonalization as a consequence. In this respect, we follow more closely the work~\cite{bardeen:1988mh} by W.~Bardeen, where however the variational principle is imposed on the expectation value of the Hamiltonian on a trial vacuum state in Hartree approximation. Note that this translates into the minimization of  the constant terms in the Hamiltonian that are produced by a normal-ordering procedure with respect to the quasiparticle operators, as also explained in~\cite{kalashnikova:2001df}. The equivalence between a diagonalization condition and a variational principle after normal-ordering is a general fact of quadratic Hamiltonians and has already been pointed out elsewhere. We conclude that our zero-point action $S_0$, arising from the terms that, after the expansion of the partition function on the new canonical coherent states, do not depend on the quasiparticle fields, is the functional equivalent of the normal-ordering constant terms in canonical approach.

\subsubsection{Chiral condensate}
The vacuum contribution we have just evaluated is the leading order term when $N_c$ is large, so it can be used to estimate the expectation value of relevant physical quantities on the ground state of the theory. In order to investigate the possible spontaneous breaking of chiral symmetry, we search for the condensate $\braket{\bar{\psi}\psi}$, with the average now intended over all the fields of the theory. Note that, from the starting partition function~\eqref{eq:transfer_partition_path}, we can simply obtain this quantity differentiating with respect to the fermion bare mass $m$ (see~\cite{caracciolo:2010rm}). At leading order
\begin{equation}
\braket{\bar{\psi}(x)\psi(x)} = -\frac{1}{V }\frac{\partial \ln \mathcal{Z}}{\partial m}  = -N_c\int \frac{\diff q}{2\pi}\cos\bar{\theta}_q .
\end{equation}
This result is well known in literature, since~\cite{li:1986gf,bardeen:1988mh}. Using the numerical solution of equation~\eqref{eq:vacuum_theta_saddle} it is easy to find, in the chiral limit,
\begin{equation}
\left.\braket{\bar{\psi}(x)\psi(x)} \right|_{m=0} = -0.29 N_c \sqrt{2\gamma}
\end{equation}
as in \cite{li:1986gf}. Analytically, from \cite{zhitnitsky:1985um}, we know that
\begin{equation}
\left.\braket{\bar{\psi}(x)\psi(x)} \right|_{m=0} = -\frac{N_c}{\sqrt{12}} \sqrt{2\gamma},
\end{equation}
in perfect agreement. The apparent contradiction with Coleman's theorem is avoided because of the large $N_c$ limit (see~\cite{schon:2000qy} for a review).

\subsection{Quasiparticle action}
\label{subsec:quasiparticle}
We will now switch on the quasiparticle contribution to the action, in order to find informations about the excited states of the theory. From the perturbative point of view we adopted when we set up the weak coupling expansion~\eqref{eq:vacuum_weak_coupling}, the fermion action can be written as a series in $g$ that, to second order, reads as
\begin{equation}
S_{QP} \simeq S_{QP}^{(0)} + g S_{QP}^{(1)} + g^2 S_{QP}^{(2)}.
\label{eq:quasi_action}
\end{equation}
This series can be calculated, as we will do in the next paragraphs, using the same arguments we formulated to obtain the zero-point action. However, the integration over gauge fields in~\eqref{eq:transfer_partition_transfer} acts on the \emph{exponential} of the action: we should first expand it to the desired order in $g$ and then averaging over gluons, eventually re-exponentiating the result. We get
\begin{multline}
\exp\left(-S_{QP}\right) \simeq \exp\left(-S_{QP}^{(0) } - g^2 S_{QP}^{(2)} \right)\\
\cdot \left[1 - g S_{QP}^{(1)} + \frac{1}{2} g^2 \left(S_{QP}^{(1)}\right)^2 \right].
\end{multline}
The term proportional to $\left(S_{QP}^{(1)}\right)^2$, which we would miss if we had not expand the exponential, is \emph{quartic} in the fields $\alpha$, $\beta$: it could be interpreted as an interaction term between effective mesons composed of quasifermions, bound by gluon exchange. We will discuss it in paragraph~\ref{subsubsec:quasi_potential}, while now we will focus on the bilinear terms in $\alpha$, $\beta$, which can be obtained mediating directly the action~\eqref{eq:quasi_action}.

\subsubsection{Mixing terms}
To begin the discussion about the quasiparticle bilinear part, we will now calculate explicitly the terms in~\eqref{eq:effective_action_qp} that mix the $\alpha$ and $\beta$ fields. By Hermiticity and commutativity we have
\begin{equation}
\mathcal{I}^{(1,2)}_t  = \mathcal{I}^{(2,1)}_{t+1} \equiv \mathcal{I}_t.
\end{equation}
With the same arguments used in the previous section, we get
\begin{multline}
\mathcal{I}^{(0)}(p,q) = 2\pi a_0 a_1 \delta(p+q)\left(- m \sin\theta_q + q \cos\theta_q\right),\\
\shoveleft{\mathcal{I}^{(1)}_t(p,q) = ia_0a_1 A_{0,t}(p+q)\sin \frac{\theta_p + \theta_q}{2},}\\
\shoveleft{\mathcal{I}^{(2)}_t(p,q) = a_1 a_0^2 \int \frac{\diff k}{2\pi}A_{0,t}(p+k)A_{0,t}(-k +q)}\\
\cdot \left[- \frac{1}{2}\sin \frac{\theta_p + \theta_q}{2} +  \cos \frac{\theta_q - \theta_k}{2}\sin \frac{\theta_p + \theta_k}{2} \right].
\label{eq:quasi_mixing_nogauge}
\end{multline}
Integrating over the gauge fields,
\begin{multline}
\braket{\mathcal{I}_t}(p,q) = 2\pi a_0 a_1 \delta(p+q)\biggl[- m \sin\theta_q + q \cos\theta_q \\
- g^2 \int \frac{\diff k}{2\pi}\frac{\Theta\cdot\Theta}{(q-k)^2} \cos \frac{\theta_q - \theta_k}{2}\sin \frac{\theta_q - \theta_k}{2} \biggr].
\end{multline}
Inserting the expression for the Casimir operator~\eqref{eq:QCD2_Casimir} and contracting the colour indices we get, for the mixing terms,
\begin{multline}
\sum_t\braket{\beta_t \mathcal{I}_t \alpha_t + \alpha^\dagger_t \mathcal{I}_t\beta^\dagger_t} \\
= \frac{a_0}{a_1}\sum_t \int \frac{\diff q}{2\pi} \,\mathcal{I}_q[\theta] \left[\beta_t(q)  \alpha_t(-q) + \alpha^\dagger_t(q) \beta^\dagger_t(-q)\right],
\end{multline}
with
\begin{multline}
\mathcal{I}_q[\theta] = - m \sin\theta_q + q \cos\theta_q \\
- \gamma \int \!\diff k\,\frac{1}{(q-k)^2} \cos \frac{\theta_q - \theta_k}{2}\sin \frac{\theta_q - \theta_k}{2} \,.
\end{multline}
Confronting this expression with equation~\eqref{eq:vacuum_energy} we can verify that
\begin{equation}
\mathcal{I}_q[\theta] = \frac{\diff}{\diff \theta_q} \omega_q^0[\theta].
\label{eq:quasi_mixing_saddle}
\end{equation}
Thus, if $\theta=\bar{\theta}$ is a solution of the gap equation, that is respects the extremality condition for the zero-point action, the mixing term in the quasiparticle action is null once the gauge fields have been integrated out.

\subsubsection{Kinetic terms}
At this point we have to perform the calculation of the quasiparticle Hamiltonians~\eqref{eq:effective_H}, a relevant aspect in order to understand how the mechanism of confinement is realized in our picture. We sketch it for the $\alpha$ field, as the $\beta$ field discussion is analogue. The corresponding term in~\eqref{eq:effective_action_qp} is
\begin{equation}
\alpha^\dagger_t (\nabla_t - \mathcal{H}_t)\alpha_{t+1} = \alpha^\dagger_t \left( \alpha_{t+1} -  \alpha_{t} \right) - \alpha^\dagger_t \mathcal{H}'_{t} \alpha_{t+1}.
\end{equation}
The first term in the last line is simply proportional to the lattice time (right) derivative~\eqref{eq:A_derivatives_free_rightleft}, so we only need to find the expression of the second term. Inverting~\eqref{eq:effective_RER} we obtain, from~\eqref{eq:effective_H'},
\begin{multline}
{\mathcal{H}'}^{(0)}(p,q) = 2\pi a_0 a_1 \delta(p+q)\left(m \cos\theta_q +  q \sin\theta_q\right),\\
\shoveleft{{\mathcal{H}'}^{(1)}_t(p,q) = ia_0 a_1 A_{0,t}(p+q)\cos \frac{\theta_p + \theta_q}{2},}\\
\shoveleft{{\mathcal{H}'}^{(2)}_t(p,q) =- \frac{a_1 a_0^2}{2}\int \frac{\diff k}{2\pi}A_{0,t}(p+k)A_{0,t}(-k +q)}
\\\cdot\biggl[\cos \frac{\theta_p + \theta_q}{2} - 2 \cos \frac{\theta_p - \theta_k}{2} \cos \frac{\theta_q - \theta_k}{2} \biggr].
\label{eq:quasienergy_nogauge}
\end{multline}
After the integration over gluon fields:
\begin{multline}
\braket{\mathcal{H}'}(p,q) = 2\pi a_0 a_1 \delta(p+q)\biggl[m \cos\theta_q +  q \sin\theta_q
 \\
 + \frac{g^2}{2}\int \!\diff k\,\frac{\Theta\cdot\Theta}{(q-k)^2}
\cos(\theta_q - \theta_k) 
\biggr].
\end{multline}
Inserting the expression for the Casimir operator~\eqref{eq:QCD2_Casimir} and contracting the colour indices we get, for the quasiparticle kinetic energy energy in the action,
\begin{equation}
\zeta \int \frac{\diff q}{2\pi} \,\omega^{QP}_q[\theta]\alpha^\dagger_t(q)  \alpha_{t+1}(-q),
\label{eq:quasienergy_alpha}
\end{equation}
with
\begin{equation}
\omega^{QP}_q[\theta] = m \cos\theta_q +  q \sin\theta_q  + \frac{\gamma}{2}\int \frac{\diff k}{2\pi}\,\frac{\cos(\theta_q - \theta_k)}{(q-k)^2}.
\end{equation}
Sending $A_0 \to -A_0$ to get $\mathring{\mathcal{E}}_t$ from $\mathcal{E}_t$, the $\beta$ term follows easily:
\begin{equation}
\zeta \int \frac{\diff q}{2\pi} \,\omega^{QP}_q\beta_{t+1} (q)  \beta^\dagger_t(-q),
\label{eq:quasienergy_beta}
\end{equation}
with the same quasiparticle energy $\omega^{QP}_q$.

It is clear that the integral in $\omega^{QP}_q$ diverges in a neighbourhood of $q$. This is not what happens for the  contribution $\omega^0$ to the vacuum energy in~\eqref{eq:vacuum_energy}, which is regular in the IR integration. The traditional~\cite{tHooft:1974pnl,bars:1977ud,li:1986gf,kalashnikova:2001df} way out of this problem is to define a principal value procedure subtracting the divergence and obtaining a regularized $\omega^{QP}_{q,R}$ with a structure similar to the one of $\omega^0$ (note that it is this regularized quasiparticle energy that is plotted in~\cite{li:1986gf,kalashnikova:2001df,jia:2017uul} and not $\omega^0$ as in our Figure~\ref{fig:omega}, despite the similarities of the pictures). However, following~\cite{schon:2000qy}, 
we interpret the bad behaviour of the fermions energy as a sign of confinement: in the low momentum phase the quasiparticles do not propagate individually because they would require an infinite energy to do so. As we can see, the quasiparticle contributions are $O(1)$ in $N_c$, while the term we have evaluated in the previous section is $O(N_c)$: that's the reason we neglected them, in the large $N_c$ expansion, when we were searching for the vacuum energy.

\subsubsection{Quartic potential}
\label{subsubsec:quasi_potential}
Let's move on now to the evaluation of the quartic terms in the effective action, as explained at the beginning of subsection~\ref{subsec:quasiparticle}. Practically, we have to square the sum of the $O(g)$ terms we found in the previous section (in equations~\eqref{eq:quasi_mixing_nogauge} and~\eqref{eq:quasienergy_nogauge}). We will get objects with a colour structure of the type
\begin{equation}
\left(\psi^A \Theta^a \psi^B \right)\left(\psi^C \Theta^b \psi^D \right),
\label{eq:mesons_bilinear}
\end{equation}
with $\psi^{A,\cdots}$ one of the $\alpha$, $\beta$, $\alpha^\dagger$, $\beta^\dagger$ fields. After the integration over gauge fields, the adjoint indices of the $\Theta$ matrices will be summed over, because the gluon propagator is proportional to $\delta_{ab}$, and we can thus rearrange the contractions between fermions using the Fierz identity~\eqref{eq:QCD2_Fierz}. Keeping only the leading part in $N_c$, we will end with
\begin{equation}
\sum_{a=1}^{N_c^2 - 1}\left(\psi^A \Theta^a \psi^B \right)\left(\psi^C \Theta^a \psi^D \right) \simeq \frac{1}{2} \left(\psi^A \psi^D\right)\left(\psi^B\psi^C \right).
\label{eq:mesons_bilinear_Fierz}
\end{equation}
To express them we can introduce effective ``mesonic'' and ``number'' fields made of fermion pairs:
\begin{equation}
\begin{aligned}
&\Gamma_t(p,q) =  \beta_t(p) \alpha_t(q)/\sqrt{N_c},\\
&\Gamma^\dagger_t(p,q) = \alpha^\dagger_t(p)\beta^\dagger_t(q)/\sqrt{N_c}, \\
&\Lambda^\alpha_t(p,q) = \alpha^\dagger_t(p)\alpha_{t+1}(q)/\sqrt{N_c},\\
&\Lambda^\beta_t(p,q) =  \beta^\dagger_t(p)\beta_{t+1}(q) /\sqrt{N_c}.
\end{aligned}
\label{eq:meson_fields}
\end{equation}
We can see that these objects are \emph{non-local} in momentum space: this is due to the fact that they do not represent elementary particles, but composite ones. Furthermore, they are \emph{colourless}: they carry no colour indices, because of the contractions in the bilinears; we say that they represent \emph{singlet states} in colour.

Now we have to find the momentum structure. A generic term~\eqref{eq:mesons_bilinear} is of the type
\begin{multline}
\frac{g^2}{2 a_1^4}\int \frac{\diff p}{2\pi}\frac{\diff p'}{2\pi}\frac{\diff q}{2\pi}\frac{\diff q'}{2\pi}\psi^A_t(-p)\mathcal{L}^{\scriptstyle \Rnum1}_t(p,-p')\psi^B_t(p')\\
\cdot\psi^C_{t'}(-q)\mathcal{L}^{\scriptstyle \Rnum2}_{t'}(q,-q')\psi^D_{t'}(q'),
\end{multline}
with $\mathcal{L}^{\scriptstyle \Rnum1,\Rnum2}_t$ one of the operators $\mathcal{I}^{(1)}_t$, $\mathcal{I}^{(1)}_{t-1}$, $\mathcal{H}^{(1)}_t$. Then
\begin{equation}
\begin{aligned}
\mathcal{L}^{\scriptstyle \Rnum1,\Rnum2}_t(p,-p') 
\in \biggl\{& ia_0a_1 A_{0,t}(p-p')\sin \frac{\theta_p - \theta_{p'}}{2},\\
&ia_0a_1 A_{0,t-1}(p-p')\sin \frac{\theta_p - \theta_{p'}}{2}, \\
& ia_0 a_1 g A_{0,t}(p-p')\cos \frac{\theta_p - \theta_{p'}}{2} \biggr\}.
\end{aligned}
\end{equation}
Calculating all the combinations, integrating over gauge fields, anticommuting the bilinears to obtain the composite fields \eqref{eq:meson_fields}, redefining the momenta integration variables $p\to -p$ when needed, we finally obtain
\begin{equation}
\frac{1}{2} g^2 \left(S_{QP}^{(1)}\right)^2  = -a_0 \sum_t \frac{1}{a^2_1}\mathcal{V}_t\left[\Gamma,\Gamma^\dagger,\Lambda^\alpha,\Lambda^\beta;\theta\right],
\end{equation}
with
\begin{multline}
\mathcal{V}_t[\Gamma,\Gamma^\dagger,\Lambda^\alpha,\Lambda^\beta;\theta] \\
\shoveleft{\,=-\frac{\gamma}{2}\frac{1}{ (2\pi)^2} \int \!\diff p \diff p' \diff q \diff q' \,\frac{\delta(p+p'+q+q')}{(p+p')^2}}\\
\shoveleft{\quad\cdot\biggl\{ \sin \frac{\theta_p + \theta_{p'}}{2} \sin \frac{\theta_q + \theta_{q'}}{2} } \\
\cdot\Bigl[\Gamma_t(p,q')\Gamma_t(q,p') + \Gamma^\dagger_t(p,q')\Gamma^\dagger_t(q,p') \\
\shoveright{ - \Lambda^{\alpha}_t(p,q')\Lambda^\beta_t(q,p') - \Lambda^{\beta}_t(p,q')\Lambda^\alpha_t(q,p') \Bigr]}\\
\shoveleft{\quad+\cos \frac{\theta_p + \theta_{p'}}{2} \cos \frac{\theta_q + \theta_{q'}}{2} }\\
\cdot\Bigl[\Gamma^\dagger_{t+1}(p,q')\Gamma_t(q,p')  + \Gamma_t(p,q')\Gamma^\dagger_{t+1}(q,p')\\
\shoveright{+\Lambda^{\alpha}_t(p,q')\Lambda^\alpha_t(q,p') + \Lambda^{\beta}_t(p,q')\Lambda^\beta_t(q,p')\Bigr]}\\
\shoveleft{\quad+2\cos \frac{\theta_p + \theta_{p'}}{2} \sin \frac{\theta_q + \theta_{q'}}{2} }\\
\cdot\Bigl[-\Gamma^\dagger_{t+1}(p,q')\Lambda^\alpha_t(q,p') + \Lambda^{\beta}_t(p,q')\Gamma^\dagger_{t+1}(q,p')\\
-\Lambda^\alpha_t(p,q')\Gamma_t(q,p')  + \Gamma_t(p,q')\Lambda^\beta_t(q,p')\Bigr]
\biggr\}.%
\end{multline}
This contribution to the action is quartic in the quasifermions and represents an \emph{interaction term}.

Thus, the quasiparticle action~\eqref{eq:effective_action_qp}, after the average over the gauge fields and evaluated at $\bar{\theta}$, is
\begin{multline}
S_{QP}\left[\alpha,\beta\right] \\
\shoveleft{\quad= - a_0\sum_t \frac{1}{a_1}\int \frac{\diff q}{2\pi}
\Bigl[ \alpha^\dagger_t(q) (\partial^{(-)}_t - \omega^{QP}_q)\alpha_{t+1}(-q) }\\
\shoveright{- \beta_{t+1}(q)(-\partial_t^{(+)} -\omega^{QP}_q)\beta^\dagger_t(-q)\Bigr]}\\
+ a_0 \sum_t \frac{1}{a^2_1}\mathcal{V}_t\left[\Gamma,\Gamma^\dagger,\Lambda^\alpha,\Lambda^\beta\right].
\label{eq:quasi_action_final}
\end{multline}

\subsection{Mesons}
\label{subsec:qcd2_mesons}

The procedure we used in the previous section to derive an effective quasiparticle action did not require further hypothesis other than weak coupling and small lattice spacing. However, the resulting action describes a fermionic model with quartic interaction, while we know since~\cite{tHooft:1974pnl} that the excitations surviving after the large $N_c$ limit are the mesonic ones, corresponding to planar diagrams. The straightforward interpretation of the above model in term of a collective mesonic theory is quite troublesome: to start with, while it is tempting to take the bilinear fields $\Gamma_t(x,y)$, $\Gamma^\dagger_t(x,y)$ as describing mesons, they are not proper complex fields, because of the nilpotency of the Grassmann fields $\alpha$, $\beta$, so a measure over them is not well-defined unless $N_c$ is strictly infinite. Moreover, the kinetic term in the action is still quadratic, and not quartic, in the qua\-si\-fer\-mions fields, so it cannot be expressed as a kinetic term for the bilinears (not to mention the fact that it is IR divergent). These problems can be overcome, in a quite natural way, using the method described in subsection~\ref{subsec:mesons}, which produces a legitimate effective action for mesons, well describing the model in the hypothesis of boson dominance. In the rest of this section we explain how to derive an action of the form~\eqref{eq:mesons_action_thooft} for the 't~Hooft model.

\subsubsection{Mesonic action}

From equation~\eqref{eq:mesons_action_thooft} and \eqref{eq:quasi_mixing_saddle}, we find that the linear terms in $\phi$, $\phi^\dagger$ are exactly null when integrated over the gauge fields and evaluated in $\bar{\theta}$, solution of the gap equation \eqref{eq:vacuum_theta_saddle} (in the following, we will omit the bar over $\theta$ for brevity). In the same conditions, the kinetic terms become
\begin{align}
&\begin{multlined}[.8\linewidth]
\braket{\Tr \frac{1}{N_c} \mathcal{H}'_t\phi^\dagger_{t+1} \phi_{t}} \\
= \frac{a_0}{a_1^2} \int \frac{\diff p}{2\pi} \frac{\diff q}{2\pi} \,\omega_q^{QP} \phi^\dagger_{t+1}(-q,p) \phi_{t}(-p,q) ,
\end{multlined}\\
&\begin{multlined}[.8\linewidth]
\braket{\Tr \frac{1}{N_c} \mathring{\mathcal{H}}'_t\phi_t \phi^\dagger_{t+1}} \\
=  \frac{a_0}{a_1^2} \int \frac{\diff p}{2\pi} \frac{\diff q}{2\pi} \,\omega_p^{QP} \phi^\dagger_{t+1}(-q,p) \phi_t(-p,q).
\end{multlined}
\end{align}
The interaction terms are
\begin{multline}
\braket{\Tr \left( - \frac{2 g^2}{N_c}  \phi^\dagger_{t+1} \mathring{\mathcal{H}}'^{(1)}_t \phi_t \mathcal{H}'^{(1)}_t \right)} \\
\shoveleft{{}\,=- \frac{2\gamma}{(2\pi)^2} \frac{a_0}{a_1^2} \int \diff p \diff q \diff k \diff u \,\frac{\delta(p+q-k-u)}{(k-q)^2} }\\
\cdot\cos \frac{\theta_{q} -\theta_k }{2} \cos \frac{\theta_p - \theta_u}{2}\, \phi^\dagger_{t+1}(p,q) \phi_t(-k,-u),  \label{eq:QCD2_H2_bardeen}
\end{multline}
\begin{multline}
\braket{\Tr \left(\frac{g^2}{N_c}\phi_t \mathcal{I}^{(1)}_t \phi_t \mathcal{I}^{(1)}_t \right)} \\
\shoveleft{{}\,=- \frac{\gamma}{(2\pi)^2} \frac{a_0}{a_1^2} \int \diff p \diff q \diff k \diff u \,\frac{\delta(p+q+k+u)}{(k+q)^2} }\\
\cdot \sin \frac{\theta_{q} + \theta_k}{2} \sin \frac{\theta_{u} + \theta_{p}}{2} \phi_t(p,q)\phi_t(k,u) \biggr] \label{eq:QCD2_I2_bardeen}
\end{multline}
and the same for the $\phi_t^\dagger \phi_t^\dagger$ term.


Now we are left with the terms obtained expanding the exponential of the action to order $1/N_c$, mediating over the gauge fields and then re-exponentiating the result. The new terms are
\begin{multline}
\frac{g^2}{2}\frac{1}{N_c} \sum_{t,t'} \Tr \left(\phi_t \mathcal{I}^{(1)}_t + \phi^\dagger_{t+1}\mathcal{I}^{(1)}_{t} \right)\\
\cdot\Tr \left(\phi_{t'} \mathcal{I}^{(1)}_{t'} + \phi^\dagger_{t'+1}\mathcal{I}^{(1)}_{t'} \right).
\end{multline}
However, these terms are null to order $g^2$, because the traces over colour act on something linear in $A_0 = A_0^a \Theta^a$ (and so traceless).

Thus, we find, from~\eqref{eq:mesons_action_thooft},
\begin{equation}
\begin{aligned}
S_M & = a_0\sum_t \frac{1}{a_1^2}\int \frac{\diff p}{2\pi} \frac{\diff q}{2\pi} \Biggl\{- \phi_t(-q,p) \partial^{(+)}_0 \phi^\dagger_t(-p,q) \\
& + \left(\omega_q + \omega_p\right) \phi^\dagger_{t+1}(-q,p) \phi_{t}(-p,q)\\
& -\frac{\gamma}{2}  \int \diff k \diff u \frac{\delta(p+q+k+u)}{(k+q)^2}\\
& \hspace{1em}\cdot\biggl[ 2\cos \frac{\theta_{q} +\theta_k }{2} \cos \frac{\theta_p + \theta_u}{2}\, \phi^\dagger_{t+1}(p,q) \phi_t(k,u) \\
&\hspace{2em}+\sin \frac{\theta_{q} + \theta_k}{2} \sin \frac{\theta_{u} + \theta_{p}}{2} \phi_t(p,q)\phi_t(k,u) \\
&\hspace{2em} +  \sin \frac{\theta_{q} + \theta_k}{2} \sin \frac{\theta_{u} + \theta_{p}}{2}
\phi^\dagger_t(p,q)\phi^\dagger_t(k,u) \biggr] \Biggr\} \,.
\end{aligned}
\label{eq:QCD2_action_dimensionless}
\end{equation}
The annoying $a_1^{-2}$ factor appears because we kept all the quantities dimensionful \emph{except} the spinor fields, which are dimensionless (in coordinate space: as equation~\eqref{eq:A_Fourier_dimension} holds, their Fourier components have dimension -1) to avoid lattice factors in solving the Berezin integrals in subsections~\ref{subsec:quasi_action} and \ref{subsec:mesons}. The dimensionless spinor fields are related to their dimensionful counterparts (respectively, on the LHS and RHS of the following relations) by
\begin{equation}
\alpha_t \longrightarrow a_1^{1/2} \alpha_t \,, \qquad \beta_t \longrightarrow a_1^{1/2} \beta_t \,,
\end{equation}
so the meson field (which in equation \eqref{eq:QCD2_action_dimensionless} has dimension -2, because of \eqref{eq:A_Fourier_dimension}) becomes
\begin{equation}
\phi_t \longrightarrow a_1 \phi_t \,.
\label{eq:QCD2_rescaling_meson}
\end{equation}
Now, after the rescaling,
\begin{equation}
\dim \phi_t(x,y) = 1  \,, \qquad \dim \phi_t(p,q) = -1 \,.
\end{equation}
Finally, evaluating the delta functions, we get the~\emph{effective meson action}
\begin{multline}
S_M = a_0 \sum_{t} \int \frac{\diff Q\diff q}{(2\pi)^2}\\
\shoveleft{\,\cdot \Biggl\{\phi_t(Q-q,q)\left[ - \partial^{(-)}_0 + \left(\omega_{Q-q} + \omega_{q}\right)\right]\phi^\dagger_{t+1}(-q,-Q + q)}\\
\shoveleft{  - \frac{\gamma}{2}\int\frac{\diff q'}{\left(q-q'\right)^2}\biggl[2\cos \frac{\theta_{q} -\theta_{q'} }{2} \cos \frac{\theta_{Q-q} - \theta_{Q-q'}}{2}}\\
\shoveright{\cdot \phi_t(Q-q,q)\phi^\dagger_{t+1}(-q',-Q+q')}\\
\shoveleft{\qquad+\sin \frac{\theta_{q} -\theta_{q'} }{2} \sin \frac{\theta_{Q-q} - \theta_{Q-q'}}{2}}\\
\shoveright{\cdot\phi_t(Q-q,q)\phi_{t}(-q',-Q+q')}\\
\shoveleft{\qquad+\sin \frac{\theta_{q} -\theta_{q'} }{2} \sin \frac{\theta_{Q-q} - \theta_{Q-q'}}{2}}\\
\cdot\phi^\dagger_t(Q-q,q)\phi^\dagger_{t}(-q',-Q+q') \biggr] \Biggr\} \,.
\label{eq:QCD2_action}
\end{multline}

See~\cite{bardeen:1988mh,kalashnikova:2001df} for a comparison from the Hamiltonian point of view.  To recover  a quadratic theory for mesons from a fermionic quartic one, in~\cite{bardeen:1988mh} it is remarked that the matrix elements  of the kinetic quasifermion Hamiltonian and the ones of a suitable quadratic term of mesons, when evaluated on mesonic states of the type~\eqref{eq:mesons_condensate}, are the same. The authors of~\cite{kalashnikova:2001df} substitute bilinear with quartic fermionic operators with the same (large-$N_c$) commutation relations, noting that in the mesonic sector single-quarks excitations are suppressed. In our approach, the bosonization of the model is obtained via the projection~\eqref{eq:mesons_partition}.

\subsubsection{Diagonalization}
The expression~\eqref{eq:QCD2_action} still cannot be interpreted as an action for complex scalar fields representing physical mesons: the fields involved are not local, and the interaction terms are not in the usual form $\phi^\dagger \phi$. We know of two ways to proceed: the first is to recognize in~\eqref{eq:QCD2_action} an action (in holomorphic representation) obtained starting from a Nambu-like Hamiltonian in the operator doublets $(\Gamma,\Gamma^\dagger)$, which can be diagonalized with a generalized Bogoliubov transformation for bilocal bosonic operators. This is the approach of~\cite{kalashnikova:2001df}: the parameters of the transformation that diagonalize the action comply with the Bars-Green equations~\cite{bars:1977ud} for the structure functions of the mesons. This procedure also gives a large-$N_c$ subleading correction for the vacuum contribution.

The other path would be to expand, directly in the functional formalism, the bilocal functions $\phi$ and $\phi^\dagger$ on a basis of two-particles states. In the free ($\gamma=0$) case the procedure is straightforward: the fields are expanded as
\begin{equation}
\begin{aligned}
&\phi_t(Q-q,q) = \sum_n \varphi^{(n)}_t(Q) \rho^{(n)}_Q(q) \,, \\
&\phi_t^\dagger(q,Q-q) = \sum_n {\varphi^{\dagger}_t}^{(n)}(Q) \rho^{*(n)}_{-Q}(q) \,,
\end{aligned}
\label{eq:QCD2_structure_basis}
\end{equation}
where the index $n$ spans the basis of two-particles states $\rho^{(n)}_Q(q)$, which satisfy the completeness relations
\begin{equation}
\begin{gathered}
\int \frac{\diff q}{2\pi} {\rho}^{*(n)}_Q(q)\rho^{(m)}_{Q}(-q) = \delta_{nm} \,,\\
\sum_{n} {\rho}^{*(n)}_Q(q)\rho^{(n)}_{Q}(p) = 2 \pi \delta(p+q) \,,
\end{gathered}
\label{eq:QCD2_structure_completeness}
\end{equation}
while $\varphi^{(n)}_t(Q)$ are the corresponding coefficients. In this way we can separate the parts of the fields that depend only from center-of-mass momenta, which ultimately will be identified with the physical mesons, from the structure part, depending also on internal momenta. Indeed, as the basis is fixed, the functional measure over $\phi$ becomes a measure for the coefficients:
\begin{equation}
\prod_{p,p'} \diff\phi_t(p,p') \to \prod_Q \prod_n \diff \varphi^{(n)}_t(Q) \,.
\end{equation}
Because of~\eqref{eq:QCD2_structure_completeness} and the fact that the structure functions do not depend on time, the temporal derivative part is already diagonal in the indices $(n)$. In order to diagonalize also the energy part, we need to impose
\begin{equation}
\lambda_Q^{(n)}\delta_{mn} = \int \frac{\diff q}{2\pi} \left(\omega_{Q-q} + \omega_{q} \right)\rho^{(n)}_Q(q) {\rho^*_{Q}}^{(m)}(-q)
\label{eq:QCD2_structure_eigen_free1}
\end{equation}
so that, in the free case,
\begin{multline}
S_M^{(\gamma=0)} = a_0 \sum_{t} \int \frac{\diff Q}{2\pi} \\
\cdot\sum_n \varphi^{(n)}_t(Q) \left[ -\partial^{(-)}_0  + \lambda_Q^{(n)}\right]  {\varphi^{\dagger(n)}_{t+1}}(-Q) \,.
\label{eq:QCD2_structure_action_diagonal}
\end{multline}
Equation~\eqref{eq:QCD2_structure_eigen_free1} is equivalent to the request that
\begin{equation}
\left(\omega_{Q-p} + \omega_{p} - \lambda_Q^{(n)} \right)\rho^{(n)}_Q(p) = 0 \,.
\label{eq:QCD2_structure_BG_free}
\end{equation}
We see that the physical mesons form a tower of states spanned by the index $(n)$.

In the general case with $\gamma\neq 0$ the action have to be diagonalized also with respect to the Nambu doublets, in order to obtain only terms $\varphi \varphi^\dagger$. The procedure is similar as before (and similar to the one in~\cite{kalashnikova:2001df}), but now~\eqref{eq:QCD2_structure_basis} mixes the coefficients $\varphi^\dagger$ and $\varphi$, given two sets of different structure functions $\rho_+^{(n)}$ and $\rho_-^{(n)}$. It turns out that, to leave a diagonal action like~\eqref{eq:QCD2_structure_action_diagonal}, the structure functions need to satisfy the Bars-Green equations, which, in this respect, generalize equation~\eqref{eq:QCD2_structure_BG_free}. The divergences we noted in subsection \ref{subsec:quasiparticle} in the quasiparticle energies are no more present in the mesonic spectrum found in this way.

\section{Conclusions and outlooks}
\label{sec:conclusion}
In the present work we have studied the 't~Hooft model on the lattice, adopting, in a relativistic setting, methods and intuitions inspired by quantum many-body systems. In particular, the confinement of the quarks has been understood as a collective phenomenon similar to the condensation of Cooper pairs in the BCS theory of superconductivity. The existence of an energy gap between the perturbative vacuum and the one where the condensation occurs has been verified fixing the parameters of a Bogoliubov transformation on the canonical fermionic operators via a variational principle, obtaining the gap equation~\eqref{eq:vacuum_theta_saddle}. Moreover, we exploited the bridge between canonical formalism and functional formulation for a quantum partition function to obtain a relativistic theory of the fluctuations around this non-perturbative vacuum. At first, we have considered quasiparticle fermionic fluctuations, obtaining the model~\eqref{eq:quasi_action_final}. Then, we have implemented an hypothesis of composite bosons dominance on the spectrum, which in our approach has been realized by a projection onto the mesons subspace, to find an effective action~\eqref{eq:QCD2_action} of these composite fields on the lattice. We have also verified that the continuum limit of this theory reproduces lots of results already known from different approaches. In this way we have checked the full consistency of method proposed, which allows to derive the properties of the effective model from the ones of the fundamental, fermionic, theory. In this way, the effective description and the fundamental one are unified in a coherent frame.

Regarding future developments, we remark here that, as it is formulated in~\cite{caracciolo:2008ag, caracciolo:2011aa}, the method we used has already been extended to include models at finite temperature and chemical potential. Considering the great interest aroused by the study of the phase diagram of gauge theories, the path we have traced in the present work naturally leads to a formulation of the 't~Hooft model at finite density, which should, in principle, reproduce the results of~\cite{li:1986gf,schon:2000qy}. Moreover, thanks to the great generality of our approach, other models of strong interaction can be studied as well, searching for non trivial predictions about ground states and low-energy excitations. Of course, in real QCD confinement is not a perturbative matter from the point of view of the configuration of the gauge fields, so we do not expect that the present reasoning can be adopted verbatim in that context. Indeed, the description of the ground state is one of the QCD long standing problem, and has been addressed in countless ways in the past (see, for a recent survey,~\cite{reinhardt:2018roz}). Still, models that mimic some aspects of strong interactions, such as~\cite{kalashnikova:2017ssy}, can be studied in the present scheme with almost no modifications.

\appendix

\section{Useful formulas}
\label{sec:A_formulas}
\subsection{Lattice operators}
\label{subsec:A_lattice_op}
For a lattice with spacing $a$, we define the free shift operators as
\begin{equation}
\bigl[T_\mu\bigr]_{xy} = \delta_{x+a\hat{\mu},y} \,, \qquad \bigl[T^\dagger_\mu\bigr]_{xy} = \delta_{x-a\hat{\mu},y} \,.
\label{eq:A_lattice_shift}
\end{equation}
Lattice right and left derivatives are
\begin{equation}
\partial_\mu^{(+)} = \frac{1}{a}(T_\mu - 1) \,, \qquad \partial_\mu^{(-)} = \frac{1}{a}(1 - T^\dagger_\mu ) \,.
\label{eq:A_derivatives_free_rightleft}
\end{equation}
Their action on a generic function is
\begin{equation}
\partial_\mu^{(\pm)} f(x) = \pm \frac{f(x\pm a\hat{\mu}) - f(x)}{a} \,.
\label{eq:A_derivatives_free_rightleft_action}
\end{equation}
A symmetric choice for the derivative is
\begin{equation}
\partial^{(\text{s})}_\mu = \frac{1}{2}(\partial_\mu^{(+)} + \partial_\mu^{(-)}) = \frac{1}{2a}(T_\mu - T^\dagger_\mu)
\label{eq:A_derivatives_free_simm}
\end{equation}
and so
\begin{equation}
\partial^{(\text{s})}_\mu f(x) = \frac{f(x+a\hat{\mu}) - f(x-a\hat{\mu})}{2a} \,.
\label{eq:A_derivatives_free_simm_action}
\end{equation}
The free Laplacian operator on the lattice is
\begin{equation}
\partial^2 = \sum_\mu \partial_\mu^{(+)}\partial_\mu^{(-)} = \sum_\mu \frac{1}{a^2}(T_\mu + T^\dagger_\mu - 2) \,.
\label{eq:A_laplacian_free}
\end{equation}
Introducing a connection, we get the covariant derivatives
\begin{subequations}
\begin{align}
&D^{(+)}_\mu(x,y) = \frac{1}{a}\left\{U_\mu(x) \bigl[T_\mu\bigr]_{xy} - 1 \right\}\,,\\
&D^{(-)}_\mu(x,y) = \frac{1}{a}\left\{1 - \bigl[T^\dagger_\mu\bigr]_{xy}U^\dagger_\mu(y) \right\} \,,\\
&D_\mu(x,y) = \frac{1}{2a}\left\{U_\mu(x) \bigl[T_\mu\bigr]_{xy} - \bigl[T^\dagger_\mu\bigr]_{xy}U^\dagger_\mu(y)\right\} \,,
\end{align}
\label{eq:A_derivatives_cov}%
\end{subequations}
where $U_\mu(x)$ is the parallel transporter in direction $\mu$. As usual,
\begin{equation}
\slashed{D}(x,y) = \sum_\mu \gamma_\mu D_\mu(x,y) \,.
\label{eq:A_derivatives_cov_slash}
\end{equation}
The covariant Laplacian is naturally
\begin{equation}
D^2 (x,y) = \sum_\mu \frac{1}{a^2}\left\{U_\mu(x)\bigl[T_\mu\bigr]_{xy} + \bigl[T^\dagger_\mu\bigr]_{xy}U^\dagger_\mu(y) - 2\right\} \,.
\label{eq:A_laplacian_cov}
\end{equation}

\subsection{Fourier representation for the lattice in 1+d dimensions}

To expand lattice functions in the Fourier basis, we introduce the momentum representation in the spatial dimensions~\footnote{We omit the accent on $f(\vec{p})$ that usually denotes the Fourier components. However, we always write explicitly the arguments in momentum space, like $\vec{p}$ or $\vec{q}$: we warn the reader that the symbol $f$ in $f(\vec{p})$ defines a different function than the one in $f(\vec{x})=f_{\vec{x}}$.}
\begin{equation}
f(\vec{p}) = a_s^d \sum_{\vec{x}} f(\vec{x}) e^{-i \vec{p}\cdot \vec{x}}, \quad f(\vec{x}) =  \int_{BZ}\frac{\Diff{d} \vec{p}}{(2\pi)^d}\, f(\vec{p}) e^{i\vec{p}\cdot \vec{x}},
\label{eq:A_Fourier}
\end{equation}
where $\vec{x}$ and $\vec{p}$ denote the spatial $d$-vectors of position and momentum ($d+1$)-vectors (which are dimensional variables with dimension of mass, respectively, -1 and 1) and $f(\vec{p})$ is a periodic function in $\vec{p}$ (it takes values in the first Brillouin Zone, $BZ = (-\pi/a_s,\pi/a_s]^d$). As $\vec{x}$ is a discrete variable, we have simply defined
\begin{equation}
\sum_\vec{x} = \sum_\vec{n} = \sum_{n_1,\cdots,n_d}\,, \qquad \vec{x} = a_s \vec{n} \,,\quad \vec{n} \in \mathbb{Z}^d \,.
\end{equation}
Therefore, the delta functions can be represented on the lattice as
\begin{equation}
\begin{aligned}
&\delta^d(\vec{p}-\vec{q}) = \left(\frac{a_s}{2\pi}\right)^d \sum_\vec{x} e^{-i\vec{x}\cdot(\vec{p}-\vec{q})} \,,\\
&\frac{\delta^d_{\vec{x}\vec{y}}}{a_s^d} = \int_{BZ} \frac{\Diff{d} \vec{p}}{(2\pi)^d}\, e^{i\vec{p}\cdot (\vec{x} - \vec{y})} \,.
\end{aligned}
\label{eq:A_Fourier_delta}
\end{equation}
In this respect we follow~\cite{rothe:lgt} (see section 2.5). These definitions tend (formally) to the usual ones for the continuum Fourier transform in the limit $a_s \to 0$.

For a matrix $A$ our convention is to use~\eqref{eq:A_Fourier} for each index:
\begin{equation}
\begin{aligned}
&A(\vec{p},\vec{q}) = a^{2d}_s \sum_{\vec{x},\vec{y}} A_{\vec{x}\vec{y}}e^{-i(\vec{p}\cdot \vec{x}+\vec{q}\cdot\vec{y})} \,,\\
&A_{\vec{x}\vec{y}} = \int_{BZ} \frac{\Diff{d} \vec{p}}{(2\pi)^d}\frac{\Diff{d} \vec{q}}{(2\pi)^d}\, A(\vec{p},\vec{q}) e^{i(\vec{p}\cdot \vec{x}+\vec{q}\cdot\vec{y})}
\end{aligned}
\end{equation}
so, by dimensional analysis,
\begin{equation}
\dim f(\vec{p}) = \dim f(\vec{x}) - d , \,\,\, \dim A(\vec{p},\vec{q}) = \dim A_{\vec{x}\vec{y}} - 2d
\label{eq:A_Fourier_dimension}
\end{equation}
where ``$\dim$'' indicates the physical dimension in mass unit. Using these formulas it's easy to proof that, for any two operators $A$, $B$ and functions $f$, $g$, the following convolution relations hold:
\begin{align}
&\tr A = \sum_\vec{x} A_{\vec{x}\vec{x}} = a_s^{-d}\!\int_{BZ} \frac{\Diff{d} \vec{p}}{(2\pi)^d}\,  A(\vec{p},-\vec{p}) \,, \label{eq:A_Fourier_trace}\\
&\!\begin{multlined}[.85\linewidth]
(AB)_{\vec{x}\vec{y}} = \int_{BZ} \frac{\Diff{d} \vec{p}}{(2\pi)^d} \frac{\Diff{d} \vec{q}}{(2\pi)^d}\, e^{i(\vec{p}\cdot \vec{x}+\vec{q}\cdot\vec{y})} \\
\cdot\left[a_s^{-d}\!\!\int_{BZ} \frac{\Diff{d} \vec{k}}{(2\pi)^d}\,  A(\vec{p},\vec{k})B(-\vec{k},\vec{q})\right], \label{eq:A_Fourier_product}
\end{multlined}
\\
&\!\begin{multlined}[.85\linewidth]
Af(\vec{x}) = \!\int_{BZ} \frac{\Diff{d} \vec{p}}{(2\pi)^d}e^{i\vec{p}\cdot\vec{ x} }\\
\cdot \left[ a_s^{-d}\!\! \int_{BZ}\frac{\Diff{d} \vec{q}}{(2\pi)^d}   A(\vec{p},\vec{q})f(-\vec{q}) \right] ,
\end{multlined}
\\
&\sum_{\vec{x}} f(\vec{x}) g(\vec{x}) = a_s^{-d}\!\int_{BZ} \frac{\Diff{d} \vec{p}}{(2\pi)^d} f(\vec{p}) g(-\vec{p}),\\
&\!\begin{multlined}[.85\linewidth]
\sum_{\vec{z}} A_{\vec{x}\vec{z}} f(\vec{z}) B_{\vec{z}\vec{y}} = \int_{BZ} \frac{\Diff{d} \vec{p}}{(2\pi)^d} \frac{\Diff{d} \vec{q}}{(2\pi)^d} e^{i(\vec{p}\cdot\vec{x}+\vec{q}\cdot\vec{y})}\\
\cdot \left[ a_s^{-d}\! \int_{BZ}  \frac{\Diff{d} \vec{k}}{(2\pi)^d}  \frac{\Diff{d} \vec{s}}{(2\pi)^d}\,  A(\vec{p},\vec{k})f(-\vec{k}+\vec{s}) B(-\vec{s},\vec{q}) \right]
\end{multlined}
\end{align}
and so on.

Conventions on complex and Hermitian conjugation are \emph{the same}:
\begin{equation}
\begin{aligned}
&f^*(\vec{x}) =  \int_{BZ}\frac{\Diff{d} \vec{p}}{(2\pi)^d}\, f^*(\vec{p}) e^{i\vec{p}\cdot \vec{x}} \,, \\
&f^*(\vec{p}) = a_s^d \sum_{\vec{x}} f^*(\vec{x}) e^{-i \vec{p}\cdot \vec{x}} \,, \\
&\left[A^\dagger\right]_{\vec{x}\vec{y}} = \int_{BZ} \frac{\Diff{d} \vec{p}}{(2\pi)^d}\frac{\Diff{d} \vec{q}}{(2\pi)^d}\, A^\dagger(\vec{p},\vec{q}) e^{i(\vec{p}\cdot \vec{x}+\vec{q}\cdot\vec{y})} \,,\\
&A^\dagger(\vec{p},\vec{q}) = a^{2d}_s \sum_{\vec{x},\vec{y}} \left[A^\dagger\right]_{\vec{x}\vec{y}}e^{-i(\vec{p}\cdot \vec{x}+\vec{q}\cdot\vec{y})} \,.
\end{aligned}
\end{equation}
Note that this means
\begin{equation}
\begin{aligned}
\left[f(\vec{p})\right]^* &= f^*(-\vec{p})\,,\\
\left[A(\vec{p},\vec{q})\right]^\dagger &= A^\dagger(-\vec{q},-\vec{p}) \,.
\end{aligned}
\label{eq:A_Fourier_conjugations}
\end{equation}

\subsection{Dirac matrices}
In 1+1 dimensions we choose the representation
\begin{equation}
\begin{gathered}
\gamma^0 = \begin{pmatrix}
1	& 0\\
0	&-1
\end{pmatrix} = \sigma^3 \,, \quad \gamma^1 = i\begin{pmatrix}
0	&	1\\
-1	&	0
\end{pmatrix} = -\sigma^2\,,\\
\gamma^5 = -i\gamma^0\gamma^1 = \begin{pmatrix}
0	&	1\\
1	&	0
\end{pmatrix} = \sigma^1 \,.
\end{gathered}
\end{equation}

\subsection{Berezin Integrals}
\label{subsec:A_Berezin}

To define a path integral for fermions it is customary to work with anticommuting (``Grassmannian'') variables. This construction is standard and can be found in virtually every textbooks about QFT. We report here the main points for practicality and to establish conventions. Let $\theta_K$, $\theta^\dagger_K$, $K=1,\cdots,\Omega$, be two (independent, in spite of the $\dagger$) sets of anticommuting symbols, that is
\begin{equation}
\{\theta_K, \theta_J \} = 0 \,, \quad \{\theta^\dagger_K, \theta^\dagger_J \}=0 \,, \quad \{\theta_K, \theta^\dagger_J \}= 0 \,.
\end{equation}
An integral over these symbols is defined by the requests
\begin{equation}
\begin{gathered}
\int\!\diff\theta_K\, 1 = 0 \,, \quad \int\!\diff\theta_K\, \theta_K = 1 \,, \\
\quad \int\!\diff\theta^\dagger_K \,1= 0 \,, \quad \int\!\diff\theta^\dagger_K\,\theta^\dagger_K =1 \,.
\end{gathered}
\end{equation}
This means that
\begin{equation}
\int\!\diff\theta_K \equiv \frac{\partial}{\partial \theta_K} \,, \qquad \int\!\diff\theta^\dagger_K \equiv \frac{\partial}{\partial \theta^\dagger_K} \,.
\end{equation}
The product measures are simply
\begin{multline}
\diff\theta \equiv \diff \theta_1 \cdots \diff\theta_\Omega, \quad \diff \theta^\dagger \equiv \diff \theta^\dagger_\Omega \cdots \diff \theta^\dagger_1 \\
 \implies \quad \diff\theta^\dagger\! \diff \theta \equiv \prod_{K=1}^\Omega \diff \theta^\dagger_K\! \diff\theta_K \,.
\end{multline}
Using these definitions, it is easy two show that the following formula for Gaussian integrals holds:
\begin{equation}
\int \diff\theta^\dagger\! \diff \theta\, e^{-\theta^\dagger A\theta + \eta^\dagger \theta + \theta^\dagger \eta} = \det A \,e^{\eta^\dagger A^{-1} \eta} \,.
\label{eq:A_Berezin_gaussian}
\end{equation}

\section{Effective action with Bogoliubov transformations}
\subsection{Canonical coherent states}
\label{subsec:A_coherent}
Let $\mathscr{F} = \bigotimes_{K=1}^\Omega \mathscr{H}_K$ be a Fock space defined as a direct product of single-particle fermionic Hilbert spaces. The annihilation and creation operators obey canonical anticommutation relations
\begin{equation}
\bigl\{ \hat{a}_J, \hat{a}_K\bigr\} = 0\,, \quad \bigl\{ \hat{a}^\dagger_J, \hat{a}^\dagger_K\bigr\} = 0\,, \quad \bigl\{ \hat{a}_J, \hat{a}^\dagger_K\bigr\} =\delta_{JK}\,,
\end{equation}
with $J,K=1,\cdots,\Omega$. For each sector, the vacuum state is defined by
\begin{equation}
\hat{a}_K\ket{0}_K = 0
\end{equation}
and therefore the vacuum state in $\mathscr{F}$ is
\begin{equation}
\ket{0} = \bigotimes_{K=1}^\Omega \ket{0}_K \,.
\end{equation}
Applying the creation operators, we obtain
\begin{equation}
\ket{K_1\cdots K_p} = \hat{a}^\dagger_{K_1}\cdots \hat{a}^\dagger_{K_p}\ket{0} \qquad p=1,\cdots,\Omega \,.
\end{equation}
These states span the entire Fock space: they form an orthonormal basis
\begin{equation}
\braket{J_1\cdots J_p| K_1\cdots K_q} = \delta_{pq}\!\sum_{\text{perm $\pi$}}\!(-)^\pi \delta_{J_1 \pi(K_1)} \cdots \delta_{J_p \pi(K_p)}
\end{equation}
and the resolution of unity reads
\begin{equation}
\sum_{p=0}^\Omega \frac{1}{p!}\sum_{K_1,\cdots,K_p} \ket{K_1\cdots K_p}\!\bra{K_1\cdots K_p} = \mathbb{I} \,.
\end{equation}
The factor $1/p!$ takes account of the indistinguishability of the particles. A generic state $\ket{\psi}$ can be written as
\begin{equation}
\ket{\psi}=\psi(\hat{a}^\dagger)\ket{0} \,, \quad \psi(\hat{a}^\dagger) \equiv \sum_{p=0}^\Omega \frac{1}{p!}\psi_{K_1\cdots K_p}\hat{a}^\dagger_{K_1}\cdots \hat{a}^\dagger_{K_p} \,,
\label{eq:A_coherent_generic_state}
\end{equation}
where the coefficients $\psi_{K_1\cdots K_p}$ are totally antisymmetric. An arbitrary operator $\hat{A}$ can be written as
\begin{equation}
\hat{A} = \sum_{p,q}\frac{1}{p!q!}A_{K_1\cdots K_p J_1 \cdots J_q} \hat{a}^\dagger_{K_1} \cdots \hat{a}^\dagger_{K_p}\hat{a}_{J_q}\cdots \hat{a}_{J_1} \,,
\label{eq:A_coherent_generic_operator}
\end{equation}
normal-ordered.

The \emph{canonical coherent states} are defined by
\begin{equation}
\ket{\theta} = e^{-\sum_K \theta_K \hat{a}^\dagger_K}\ket{0} = \prod_K \left(1-\theta_K \hat{a}^\dagger_K\right)\ket{0} \,,
\end{equation}
with $\theta_K$ Grassmannian variables such that
\begin{equation}
\begin{gathered}
\theta_J \theta_K = - \theta_K \theta_J \,, \quad \theta_J^\dagger \theta_K = - \theta_K \theta_J^\dagger \,, \\
\theta_J \hat{a}_K = - \hat{a}_K \theta_J \,, \quad \theta_J \hat{a}^\dagger_K = - \hat{a}^\dagger_K \theta_J
\end{gathered}
\end{equation}
(the variables $\theta_K$ and $\theta^\dagger_K$ are independent for the purpose of constructing the algebra).
The state $\ket{\theta}$ is an eigenstate of $\hat{a}_K$, with eigenvalues $\theta_K$:
\begin{equation}
\hat{a}_K \ket{\theta} = \theta_K \ket{\theta} \,.
\label{eq:A_coherent_eigenstate}
\end{equation}
The inner product between two coherent states $\ket{\theta}$ and $\ket{\eta}$ is
\begin{equation}
\braket{\theta|\eta} = \exp{\left(\sum_K \theta^\dagger_K \eta_K\right)} \equiv \exp\left(\theta^\dagger\eta\right) \,.
\end{equation}
These states form an overcomplete basis, and the resolution of unity reads as
\begin{equation}
\hat{\mathbb{I}}= \int\!\diff\theta^\dagger\! \diff \theta \, \frac{\ket{\theta}\!\bra{\theta}}{\braket{\theta|\theta}} \,.
\label{eq:A_coherent_unity}
\end{equation}
Using~\eqref{eq:A_coherent_unity} a generic state can be written as
\begin{equation}
\ket{\psi} = \int\!\diff\theta^\dagger\! \diff \theta \, e^{-\theta^\dagger \theta}\psi(\theta^\dagger)\ket{\theta} \,,
\end{equation}
with
\begin{equation}
\psi(\theta^\dagger) \equiv \braket{\theta|\psi} = \sum_{p=0}^\Omega \frac{1}{p!}\psi_{K_1\cdots K_p}\theta^\dagger_{K_1}\cdots \theta^\dagger_{K_p} \,,
\end{equation}
as it results from~\eqref{eq:A_coherent_generic_state} and~\eqref{eq:A_coherent_eigenstate}. Using~\eqref{eq:A_coherent_generic_operator}, the expression for the representative of a generic operator follows:
\begin{multline}
A(\theta^\dagger,\eta) \equiv \braket{\theta|\hat{A}|\eta} \\
= e^{\theta^\dagger\eta} \sum_{p,q}\frac{1}{p!q!}A_{J_1\cdots J_p K_1 \cdots K_q} \theta^\dagger_{J_1} \cdots \theta^\dagger_{J_p}\eta_{K_q}\cdots \eta_{K_1}
\end{multline}
and so
\begin{align}
&A\psi(\theta^\dagger) \equiv \braket{\theta|\hat{A}|\psi} = \int\!\diff\eta^\dagger\! \diff \eta \, e^{-\eta^\dagger \eta} A(\theta^\dagger,\eta)\psi(\eta^\dagger) \,,\\
&AB(\theta^\dagger,\eta) \equiv \braket{\theta|\hat{A}\hat{B}|\eta} \notag\\
&\qquad= \int\!\diff{\eta'}^\dagger\! \diff \eta' \, e^{-{\eta'}^\dagger \eta'} A(\theta^\dagger,\eta')B({\eta'}^\dagger,\eta) \,.
\label{eq:A_coherent_prod}
\end{align}
If an operator $\hat{O}$ can be written in the special form
\begin{equation}
\hat{O} = \exp\Bigl(\sum_{J,K}\hat{a}^\dagger_J M_{JK}\hat{a}_K\Bigr) \,,
\end{equation}
then its representative is
\begin{equation}
O(\theta^\dagger,\eta) \equiv \braket{\theta|\hat{O}|\eta} = \exp\Bigl[\sum_{J,K}\theta^\dagger_J (e^M)_{JK}\eta_K\Bigr] \,.
\label{eq:A_coherent_exp}
\end{equation}
Combining this formula with~\eqref{eq:A_coherent_prod} it follows that, if two operators are in the form
\begin{equation}
\hat{O}_1 = \exp\Bigl(\sum_{J,K}\hat{a}^\dagger_J M_{JK}\hat{a}_K\Bigr), \, \hat{O}_2 = \exp\Bigl(\sum_{J,K}\hat{a}^\dagger_J N_{JK}\hat{a}_K\Bigr),
\end{equation}
then
\begin{equation}
O_1O_2(\theta^\dagger, \eta) = \exp \Bigl[\sum_{J,K}\theta^\dagger_J (e^Me^N)_{JK}\eta_K\Bigr] \,.
\label{eq:A_coherent_exp2}
\end{equation}
Finally, the trace of an even operator $\hat{A}$ (such as it commutes with any Grassmannian variable, $\hat{A}\theta_K = \theta_K\hat{A}$) can be written as
\begin{equation}
\tr \hat{A} = \int\!\diff\theta^\dagger\! \diff \theta \, e^{-\theta^\dagger \theta} A(\theta^\dagger,-\theta) \,.
\label{eq:A_coherent_trace}
\end{equation}

If two kind of fermions (namely, particles and antiparticles) are admitted, the Fock space can be constructed from the set of canonical creation and annihilation operators
\begin{equation}
\begin{aligned}
&\bigl\{\hat{u}^\dagger_J,\hat{u}_K \bigr\} = \bigl\{\hat{v}^\dagger_J,\hat{v}_K \bigr\} = \delta_{JK}, \\
&\bigl\{\hat{u}_J,\hat{u}_K \bigr\} = \bigl\{\hat{v}_J,\hat{v}_K \bigr\} = \cdots =0,
\end{aligned}
\end{equation}
where $u^\dagger_J$ and $v^\dagger_J$ create, respectively, a particle and an antiparticle in the state $J$. Canonical coherent states can be defined in the same way:
\begin{equation}
\ket{\rho \sigma} = \exp\left(-\sum_K \rho_K \hat{u}^\dagger_K - \sum_K \sigma_K \hat{v}^\dagger_K\right)\ket{0}
\label{eq:A_coherent_state_2}
\end{equation}
and the resolution of unity reads as
\begin{multline}
\hat{\mathbb{I}}= \int\!\prod_K \diff \rho^\dagger_K\! \diff \rho_K\!\diff \sigma^\dagger_K\! \diff \sigma_K \\
\cdot e^{-\sum_J \rho^\dagger_J \rho_J-\sum_J \sigma^\dagger_J \sigma_J} \ket{\rho \sigma}\!\bra{\rho \sigma} \,.
\label{eq:A_coherent_unity_2}
\end{multline}

\subsection{More on Bogoliubov transformations}
\label{subsec:appendixBogoliubov}

We collect here some properties of the Bogoliubov transformation~\eqref{eq:unitary_bogoliubov}. With the definitions~\eqref{eq:unitary_R}, it follows that
\begin{equation}
f(\mathcal{R})\mathcal{F}^\dagger = \mathcal{F}^\dagger f(\mathring{\mathcal{R}})\,,  \qquad
f(\mathring{\mathcal{R}})\mathcal{F} = \mathcal{F} f(\mathcal{R})
\end{equation}
for any function $f$. The transformation inverse of~\eqref{eq:unitary_bogoliubov} is
\begin{equation}
\begin{aligned}
&\hat{u} = \mathcal{R}^{1/2}\left(\hat{a} + \mathcal{F}^\dagger \hat{b}^\dagger\right) \, ,\\
&\hat{v} = \left(\hat{b} -  \hat{a}^\dagger \mathcal{F}^\dagger\right) \mathring{\mathcal{R}}^{1/2} \,,
\end{aligned}
\label{eq:unitary_bogoliubov_inverse}
\end{equation}
as expected sending $\mathcal{F} \to - \mathcal{F}$. The vacuum state~\eqref{eq:unitary_vacuum} is normalized as
\begin{equation}
\braket{\mathcal{F}|\mathcal{F}} = \det \left(\mathcal{R}^{-1} \right) = \left(\det \mathcal{R}\right)^{-1} \,.
\label{eq:unitary_F_norm}
\end{equation}
The norm of the new coherent states~\eqref{eq:unitary_coherent_new} is
\begin{equation}
\braket{\alpha\beta;\mathcal{F}|\alpha\beta;\mathcal{F}} = e^{\alpha^\dagger\alpha + \beta^\dagger\beta}\left(\det\mathcal{R}\right)^{-1} \,.
\label{eq:unitary_coherent_norm}
\end{equation}
The inner product with the old coherent states is
\begin{equation}
\braket{\rho\sigma|\alpha\beta;\mathcal{F}} = e^{\rho^\dagger\mathcal{F}^\dagger \sigma^\dagger - \tilde{\alpha}\rho^\dagger - \tilde{\beta}\sigma^\dagger - \beta\mathcal{F}\alpha} \,.
\label{eq:unitary_inner_rsab}
\end{equation}

\subsection{Quasiparticle action}
\label{subsec:quasi_action}
To find the trace of the transfer matrix in~\eqref{eq:effective_partition} we can write
\begin{multline}
\braket{\alpha_t\beta_t;\mathcal{F}_t | \hat{T}_t^\dagger \hat{V}_t \hat{T}_{t+1} | \alpha_{t+1}\beta_{t+1};\mathcal{F}_{t+1}} \\
= \braket{\alpha_t\beta_t;\mathcal{F}_t | \hat{T}_t^\dagger \hat{\mathbb{I}} \hat{V}_t \hat{\mathbb{I}}\hat{T}_{t+1} | \alpha_{t+1}\beta_{t+1};\mathcal{F}_{t+1}} \,,
\label{eq:effective_transfer_elements_identity}
\end{multline}
where $\hat{\mathbb{I}}$ is given by~\eqref{eq:A_coherent_unity_2}. We get
\begin{multline}
\braket{\rho\sigma| \hat{T}_t|\alpha_t \beta_t;\mathcal{F}_t} 
= \det \mathcal{F}^\dagger_{\mathcal{N},t} \\
\shoveleft{\quad\cdot
\exp\Bigl[ - \beta_t\mathcal{F}_t\alpha_t  
+ \tilde{\beta}_t\mathcal{N}_t\left(\mathcal{F}^\dagger_{\mathcal{N},t}\right)^{-1} \tilde{\alpha}_t }\\
- \tilde{\beta}_t\left(\mathring{\mathcal{F}}^\dagger_{\mathcal{N},t}\right)^{-1} e^{-\mathcal{M}_t} \sigma^\dagger 
+ \rho^\dagger e^{-\mathcal{M}_t}\left(\mathcal{F}^\dagger_{\mathcal{N},t}\right)^{-1} \tilde{\alpha}_t \\
+ \rho^\dagger e^{-\mathcal{M}_t} \left(\mathcal{F}^\dagger_{\mathcal{N},t}\right)^{-1} \mathcal{F}^\dagger_t e^{-\mathcal{M}_t} \sigma^\dagger
\Bigr] \,.
\end{multline}
In the same way
\begin{multline}
\braket{\omega\varphi| \hat{V}_t \hat{T}_{t+1} | \alpha_{t+1} \beta_{t+1};\mathcal{F}_{t+1}}\\
\shoveleft{\quad = \det \mathcal{F}^\dagger_{\mathcal{N},t+1} \exp\Bigl[ - \beta_{t+1}\mathcal{F}_{t+1}\alpha_{t+1}  }\\
+ \tilde{\beta}_{t+1}\mathcal{N}_{t+1}\left(\mathcal{F}^\dagger_{\mathcal{N},t+1}\right)^{-1} \tilde{\alpha}_{t+1}\\
 - \tilde{\beta}_{t+1}\left(\mathring{\mathcal{F}}^\dagger_{\mathcal{N},t+1}\right)^{-1} e^{-\mathcal{M}_{t+1}}\mathcal{U}_{0,t}^\dagger \varphi^\dagger \\
  +  \omega^\dagger \mathcal{U}_{0,t} e^{-\mathcal{M}_{t+1}}\left(\mathcal{F}^\dagger_{\mathcal{N},t+1}\right)^{-1} \tilde{\alpha}_{t+1}\\
 + \omega^\dagger \mathcal{U}_{0,t} e^{-\mathcal{M}_{t+1}}\left(\mathcal{F}^\dagger_{\mathcal{N},t+1}\right)^{-1} \mathcal{F}^\dagger_{t+1} e^{-\mathcal{M}_{t+1}} \mathcal{U}^\dagger_{0,t}\varphi^\dagger \Bigr] \,.
\end{multline}
The last effort:
\begin{multline}
\braket{\alpha_t \beta_t;\mathcal{F}_t | \hat{T}_t^\dagger \hat{V}_t \hat{T}_{t+1} | \alpha_{t+1} \beta_{t+1};\mathcal{F}_{t+1}}\\
\shoveleft{\quad=  \det \left(e^{-\mathcal{M}^\dagger_t} \mathcal{U}_{0,t} e^{-\mathcal{M}_{t+1}} \mathcal{E}_{t+1,t}\right)\exp\Bigl(
\alpha^\dagger_t \mathcal{I}^{(1,2)}_t \beta^\dagger_t}
\\
+ \beta_{t+1}\mathcal{I}^{(2,1)}_{t+1}\alpha_{t+1} + \alpha_t^\dagger \mathcal{R}^{-1/2}_{t}\mathcal{E}_{t+1,t}^{-1} \mathcal{R}^{-1/2}_{t+1}\alpha_{t+1}
\\
- \beta_{t+1}\mathring{\mathcal{R}}^{-1/2}_{t+1}\mathring{\mathcal{E}}_{t+1,t}^{-1}\mathring{\mathcal{R}}^{-1/2}_{t}\beta^\dagger_t\Bigr) \,,
\end{multline}
where we used the definitions~\eqref{eq:effective_E} and~\eqref{eq:effective_I}.

\bibliography{biblio}

\end{document}